\documentclass[12pt]{article}
\pdfoutput=1
\usepackage{draft}
\usepackage{comment}
\usepackage[normalem]{ulem}
\usepackage{hyperref}
\hypersetup{
 colorlinks=true,
 urlcolor=blue,
 anchorcolor=blue,
 citecolor=blue,
 filecolor=blue,
 linkcolor=blue,
 menucolor=blue,
 pagecolor=blue,
 linktocpage=true,
}
\usepackage{graphicx,color,subfig}
\usepackage{cite}
\usepackage{empheq} 
\usepackage[font=small,labelfont=bf]{caption} 

\usepackage[usenames,dvipsnames,table]{xcolor}
\graphicspath{{./figures/}}
\usepackage{tikz}
\usepackage{empheq}

\def\rar{\rightarrow}
\def\t{\tau}
\def\o{\over}
\def\i{\infty}
\def\foot{\footnote}
\newcommand{\es}[2] {\begin{equation} \label{#1} \begin{split} #2 \end{split} \end{equation}}
\newcommand{\e}[2] {\begin{equation} \label{#1} #2 \end{equation}}
\def\eqr{\eqref}
\def\hb{\overline{h}}
\def\g{\gamma}

\newcommand{\ZZ}{\mathbb{Z}}
\newcommand{\RR}{\mathbb{R}}
\newcommand{\HH}{\mathbb{H}}

\DeclareMathOperator*{\Res}{Res}

\DeclareMathOperator{\vol}{vol}

\newcommand{\half}{{1\over 2}}

\DeclareMathOperator{\im}{Im}
\DeclareMathOperator{\re}{Re}

\DeclareMathOperator{\diag}{diag}

\DeclareMathOperator{\rnrn}{RN}

\usepackage{scalerel}
\usepackage{stackengine,wasysym}

\DeclareFontFamily{OT1}{pzc}{}
\DeclareFontShape{OT1}{pzc}{m}{it}{<-> s * [1.10] pzcmi7t}{}
\DeclareMathAlphabet{\mathpzc}{OT1}{pzc}{m}{it}

\definecolor{vert}{rgb}{0.1367 0.543 0.1367}

\def\({\left(}
\def\){\right)}
\newcommand{\nn}{\nonumber}

\begin{document}

\unitlength = .8mm

\begin{titlepage}

	  \begin{flushright}
	 \end{flushright} 

\begin{center}

 \hfill \\
 \hfill \\

\title{Harmonic analysis of 2d CFT partition functions}

~\vskip 0.01 in

\author{
	Nathan Benjamin$^{a}$, Scott Collier$^{a}$, A. Liam Fitzpatrick$^{b}$, \\ Alexander Maloney$^{c}$, Eric Perlmutter$^{d,e}$
}
%
${}^a$\emph{\small Princeton Center for Theoretical Science, Princeton University, Princeton, NJ 08544, USA}
\\
${}^b$\emph{\small Department of Physics, Boston University, Boston, MA 02215, USA}
\\
${}^c$\emph{\small Department of Physics, McGill University, Montreal, QC H3A 2T8, Canada}
\\
${}^d$\emph{\small Institut de Physique Th\'eorique, CEA Saclay, CNRS, 91191 Gif-sur-Yvette, France}
\\
${}^e$\emph{\small Walter Burke Institute for Theoretical Physics, Caltech, Pasadena, CA 91125, USA}
~\vskip .2 in
\email{
	nathanb@princeton.edu, scott.collier@princeton.edu, fitzpatr@bu.edu, maloney@physics.mcgill.ca, perl@ipht.fr
}

\end{center}

\abstract{
We apply the theory of harmonic analysis on the fundamental domain of $SL(2,\ZZ)$ to partition functions of two-dimensional conformal field theories. We decompose the partition function of $c$ free bosons on a Narain lattice into eigenfunctions of the Laplacian of worldsheet moduli space $\mathbb H/SL(2,\mathbb Z)$, and of target space moduli space $O(c,c;\mathbb Z)\backslash O(c,c;\mathbb R)/O(c)\times O(c)$. This decomposition manifests certain properties of Narain theories and ensemble averages thereof. We extend the application of spectral theory to partition functions of general two-dimensional conformal field theories, and explore its meaning in connection to AdS$_3$ gravity. An implication of harmonic analysis is that the local operator spectrum is fully determined by a certain subset of degeneracies. 
}

\vfill

\end{titlepage}

\eject

\begingroup

\baselineskip .168 in
\tableofcontents

\endgroup

\section{Introduction and summary: can one hear the shape of a CFT?}
\label{sec:intro} 

It is often illuminating to ask how a quantum field theory behaves when it is deformed. When the theory is a conformal field theory (CFT), a particularly useful concept of deformations is the ``space of CFTs,'' on which an individual theory is just a single point.  There are two natural and common ways to formulate the space of CFTs, both of which lead to a rich and fascinating structure.  The first is to use the fact that conformal symmetry (or some extension of it) together with the operator product expansion (OPE) allow one to define CFTs in terms of a discrete set of ``data'' -- typically, operator dimensions and OPE coefficients -- so that, roughly speaking, the space of CFTs is the space of such data.  One can then study which points in this space satisfy abstract CFT axioms.  Remarkably, such constraints can lead not only to the exclusion of large regions of CFT data, but can sometimes sharply pinpoint specific theories living at the boundary of the allowed region, and can even be a practical method for computing results about a specific universality class of CFTs.  The challenge in this conformal bootstrap approach is to extract the constraints in an efficient and useful way.   The second concept of a space of CFTs arises when a theory contains one or more exactly marginal operators, which connect continuous families of CFTs. Understanding this moduli space of deformations can provide a new perspective on the original CFT.  

Our goal in this paper is to describe a  set of tools -- and more generally, an organizing principle -- for 2d CFTs that sheds light on both of these approaches. The main technology will be harmonic analysis on the fundamental domain $\mathcal{F} = \mathbb H/SL(2,\mathbb Z)$.

In two dimensions, CFTs have a rich structure due to the infinite-dimensional conformal symmetry as well as modular covariance of local torus observables. For example, the torus partition function must be modular-invariant. Viewing the partition function as a positive sum of characters of the Virasoro algebra, the mechanism by which modular invariance is satisfied remains obscure and rather remarkable. On the other hand, harmonic analysis on $\mathcal{F}$ builds in the constraint of modular invariance automatically. This comes at the expense of obscuring unitarity and discreteness, in a reversal of the typical modular bootstrap method \cite{Hellerman:2009bu}.\footnote{Harmonic analysis on the Euclidean conformal group \cite{Dobrev:1977qv} has been an extremely useful tool for the study of CFT four-point functions \cite{Cornalba:2006xm,Costa:2012cb,Cornalba:2007fs}. It has played a central role in numerous recent developments including the Lorentzian inversion formula and the SYK model \cite{Maldacena:2016hyu, Gadde:2017sjg, Hogervorst:2017sfd,Caron-Huot:2017vep,Simmons-Duffin:2017nub,Kravchuk:2018htv,Karateev:2018oml,Liu:2018jhs}. There are also conceptual similarities between our approach and the Polyakov-Mellin bootstrap for four-point functions \cite{Polyakov:1974gs,Gopakumar:2016wkt,Gopakumar:2016cpb,Gopakumar:2018xqi,gopak}, where crossing symmetry is completely manifest but positivity and analyticity properties are obscured.} Fortunately, mathematicians have extensively developed the theory of harmonic analysis on $\mathcal{F}$, reviewed in e.g. \cite{Terras_2013}. All square-integrable modular functions can be uniquely decomposed into a complete set of eigenfunctions of the Laplacian on the upper half-plane. These basis functions, comprised of both a continuous and discrete series, are themselves fully modular-invariant. 

In promoting the role of this technology in 2d CFT, we will apply it to the Narain family of free boson CFTs, and then to general 2d CFTs. Let us now discuss these in turn.

The Narain lattice CFTs, well-known to string theorists from the literature on toroidal compactifications, turn out to be especially well-suited to the application of spectral methods because their central charge is saturated by currents generating the $U(1)^c \times U(1)^c$ algebra. Spectral decomposition of these CFT partition functions -- more precisely, the $U(1)^c \times U(1)^c$ primary-counting partition functions (henceforth ``primary partition functions'') -- makes certain properties of the theories manifest. One of the main advantages of the spectral representation is that it naturally separates the partition function into a piece that is constant over the moduli space, and residual pieces whose averages on the moduli space vanish. Consequently, the recently-studied ensemble averages \cite{Maloney:2020nni, Afkhami-Jeddi:2020ezh} become completely transparent. The residual pieces are square-integrable and hence admit a spectral decomposition, while the moduli-independent piece, though ``almost'' but not quite square-integrable, can be treated using techniques of \cite{zbMATH03796039}. Moreover, in many cases we can obtain the explicit spectral decomposition of the residual pieces, which provide a concrete form for the deviation from the ensemble average. 

One of our most surprising findings is that, at least in the cases considered herein, the Narain lattice primary partition functions have a simple overlap with all of the discrete eigenfunctions, known as Maass  cusp forms.\footnote{These are in contrast with the continuous, plane-wave normalizable eigenfunctions, which are real analytic Eisenstein series.}  What makes this result surprising is that the cusp forms themselves cannot be written in closed form and are, in a precise mathematical sense that we will review \cite{sarnak, PhysRevLett.69.2188,1993MaCom..61..245H,Steil:1994ue, Sarnak_1987}, {\it chaotic} linear combinations of elementary functions. For the special case of the $c=1$ free boson compactified on a circle of radius $r$, we show that the overlap of the partition function with the cusp forms vanishes for any $r$. The same is true for $c<1$ minimal models.\footnote{This statement is slightly subtle. In all cases we decompose the CFT partition function after first dividing by powers of the Dedekind $\eta(\t)$ function, dressed by factors of $\im\tau$ to retain modular invariance. For Narain lattices, there is one obvious natural choice for the power of $\eta(\t)$, but for minimal models there are multiple natural choices and the statement only holds true for one of them.}  However, while one might reasonably have expected the same to be true of all $c$, we show that at $c=2$, as well as at $c>2$ in regions of the moduli space where we can perform the computation, the overlap with the cusp forms does not vanish but instead has an essentially closed form: in a sense to be made precise, the resulting overlap is equal to the number 8 for all cusp forms. This closed-form relation between the spectrum of free bosons on a Narain lattice, which is an integrable model, and cusp forms, which exhibit chaotic properties, deserves further study.

Certain other properties are nicely encoded in the spectral decomposition of Narain lattice partition functions.\foot{For one, this decomposition was used in \cite{Angelantonj:2011br} as an elegant way to regulate one-loop string theory integrals while manifestly preserving worldsheet modular invariance.} The fact that the Narain lattice partition functions are eigenfunctions of the difference between the target space and worldsheet Laplacians \cite{Obers:1999um,Maloney:2020nni} arises in the spectral decomposition as the fact that the coefficients of the eigenfunctions of the worldsheet Laplacian are target space Laplacian eigenfunctions with a correlated eigenvalue.  In the special case of $c=2$, the ``triality'' under exchange among the three complex moduli \cite{Dijkgraaf:1987jt}, that of the worldsheet and two from the target space, is made completely manifest in the spectral representation.  Perhaps most interesting from a modular bootstrap perspective is that the scalar primary spectrum in Narain CFTs is completely determined by the overlaps with the continuous (rather than discrete) series of basis eigenfunctions in the spectral decomposition, and therefore it provides a new handle on the problem of determining the maximal gap in the scalar primary spectrum in the Narain lattice moduli space. We demonstrate this explicitly at $c=2$.

Turning now to ``generic'' 2d CFTs, by which we mean those with Virasoro symmetry alone and $c>1$, spectral decomposition remains a powerful, if somewhat trickier, technique to implement. Harmonic analysis is applicable to modular objects that are square-integrable -- or close to it, in a sense articulated by Zagier \cite{zbMATH03796039} and used heavily here -- but primary partition functions of generic CFTs do not obey this constraint: instead, they possess exponential divergences at the cusp at $\t\rar i\infty$, due to the vacuum state and any other ``light'' primaries, defined as those with conformal dimension $\Delta \leq {c-1\o 12}$. Their presence forces us to address the key question of how to massage partition functions into a form fit for spectral decomposition, and what this means physically. We are led to advance the following perspective for general 2d CFTs. The spectral decomposition may be applied to a partition function after subtracting off all light states in a modular-invariant way. We refer to this as subtracting the ``modular completion'' of the light spectrum. The remainder after subtraction, which we call $Z_{\rm spec}$, is modular-invariant and square-integrable, and thus admits a spectral decomposition. 

\begin{figure}[t]
\centering
\includegraphics[scale=.6]{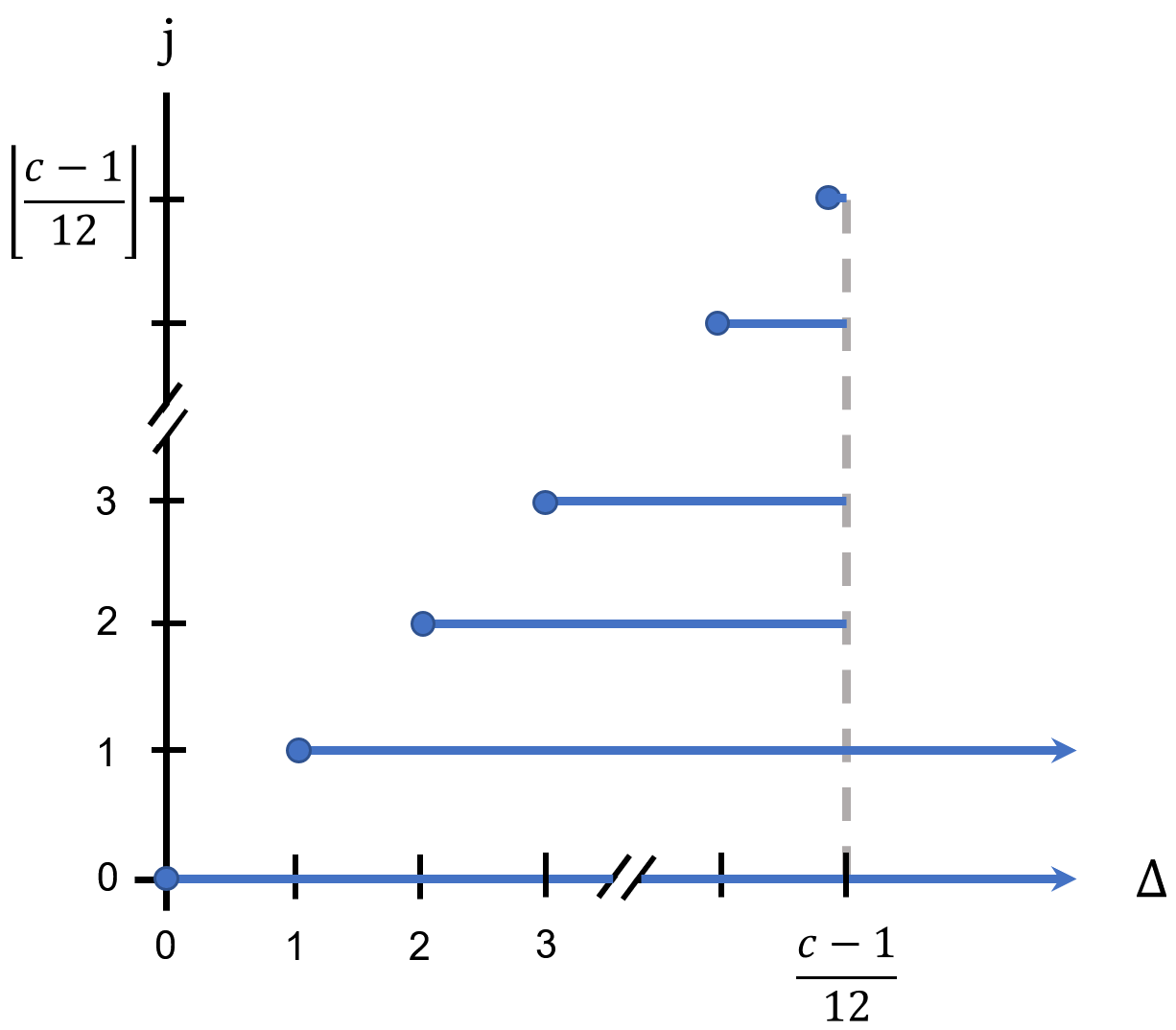}
\caption{In a 2d CFT, the full Virasoro primary spectrum is determined by the primary spectrum in blue. Circles represent the unitarity bound, $\Delta\geq j$, where $\Delta=h+\hb$ and $j=|h-\hb|$. This statement assumes that the cuspidal eigenspectrum is non-degenerate, an unproven but widely held property of $SL(2,\mathbb{Z})$. The $j=1$ data may, subject to a further mild assumption about Maass cusp forms, be replaced by the data of any fixed integer spin $j>0$ without affecting this conclusion. This result is explained in Section \ref{secspecdet}.}
\label{specfig}
\end{figure}

One rigorous consequence of this structure that emerges rather easily, as shown in Section \ref{secspecdet}, is a new result on {\it spectral determinacy}, i.e. the minimal content necessary to fully determine the spectrum of a 2d CFT \cite{Douglas:2010ic,Kaidi:2020ecu}. We show that, under the widely-held assumption that the cuspidal eigenspectrum of $SL(2,\ZZ)$ is non-degenerate (e.g. \cite{cusp82,hejh,sque}), the entire primary spectrum of a 2d CFT is {\it uniquely} fixed by the light spectrum, the scalar spectrum, and the spectrum of any single nonzero integer spin $j$ (see Fig.~\ref{specfig}).\footnote{Strictly speaking, this statement can be proven only for spin $j=1$.  For $j>1$ this relies on the additional conjecture that the Fourier coefficients of cusp forms are all non-zero. As we describe in Section \ref{secspecdet} this conjecture is almost certainly true, but unproven.} As we will discuss, we think it quite plausible that this result could be strengthened to use even less data as input. 

Physically, the modular completion of light states represents a  universal part of the CFT, or perhaps a kind of ``average,'' implied by the existence of the light spectrum. The remainder, $Z_{\rm spec}$, captures the deviation of a given CFT spectrum around this universal part. That spectral analysis of CFT partition functions is sensitive to the distinction between light and heavy operators jibes with central properties of the black hole spectrum of quantum gravity in AdS$_3$. We explain how this interpretation provides a 2d CFT analog of the ``half-wormholes'' of \cite{Saad:2021rcu}. This viewpoint has some limitations, and we discuss some concrete avenues for further research that could strengthen it. Fortunately, it is buttressed by application to the Narain case, in the following way. In situations where the partition function of interest belongs to an exact moduli space of CFTs, the sense of ``average'' used above may be given a literal interpretation. As noted earlier, the partition function $Z^{(c)}$ of a fixed Narain lattice CFT is a sum of the square-integrable terms, which average to zero, and the ensemble average with respect to the Zamolodchikov metric, $\langle Z^{(c)} \rangle$; in harmony with the above paradigm, the known result \cite{Afkhami-Jeddi:2020ezh,Maloney:2020nni} for the average $\langle Z^{(c)} \rangle$ turns out to be precisely equal to the modular completion, via Poincar\'e sum, of the vacuum state.

This paper is organized as follows. In Section \ref{sec:harmonicAnalysis}, we introduce harmonic analysis on the fundamental domain $\mathcal{F} = \mathbb{H}/SL(2,\mathbb{Z})$. In Section \ref{sec:narain}, we apply our technology to Narain’s family of free boson CFTs. In Section \ref{sec4}, we apply our technology to general 2d CFTs.

\emph{A word on notation:} In this paper all functions of complex variables are assumed to be non-holomorphic. However, for the sake of brevity of notation, we will drop all anti-holomorphic coordinate dependence. We will write e.g. $f(z)$ instead of $f(z, \bar{z})$, but $f$ should not be assumed to be a holomorphic function. We will also denote the real and imaginary parts of the torus modular parameter $\tau$ as $x$ and $y$, respectively.

\section{Harmonic analysis on the fundamental domain}\label{sec:harmonicAnalysis}
We begin by collecting some basic properties of the spectral theory of the Laplacian on the fundamental domain,
\begin{equation}
	\mathcal{F} = \HH/SL(2,\ZZ) = \left\{\tau = x + i y \in \HH \,\bigg|\, -\half < x \leq \half,~ |\tau| \geq 1\right\}.
\end{equation}
Many important results of this subject are nicely summarized in Chapter 3 of \cite{Terras_2013}. The Laplacian on the upper half-plane $\HH$ is given by
\begin{equation}
	\Delta_\tau = -y^2\left(\partial_x^2+\partial_y^2\right).
\end{equation}
Throughout this paper we will make use of the Petersson inner-product on the space $L^2(\mathcal{F})$ of square-integrable modular-invariant functions,
\begin{equation}
	(f,g) \coloneqq \int_{\mathcal{F}}{dxdy\over y^2}f(\tau) \overline{g(\tau)}.
\end{equation}

\subsection{Spectral resolution of the Laplacian}
The spectrum of the Laplacian on the fundamental domain includes both discrete and continuous components. The discrete part can be expanded in an orthogonal basis\footnote{It will turn out to be convenient to use a basis of cusp forms that are not unit-normalized. See Appendix \ref{subApp:cusp} for more details.} of \textbf{Maass cusp forms},
\begin{equation}
\begin{aligned}
	\nu_0 =&  \, \sqrt{3\over \pi} = \vol(\mathcal{F})^{-\half}\\
	\{\nu_{n\ge 1}\}:& ~~ \Delta_\tau \nu_n(\tau) = \left({1\over 4} + R_n^2\right)\nu_n(\tau), ~ R_n > 0
\end{aligned}
\end{equation}
The lowest one, $\nu_0$, is a constant. The cusp forms are distinguished by having no scalar component\footnote{In what follows unless stated otherwise the label $n$ will denote cusp forms with $n \ge 1$.}
\begin{equation}
	\int_{-\half}^{\half}dx\, \nu_n(\tau) = 0
\end{equation}
and by the fact that they decay exponentially at the cusp,
\begin{equation}
	\nu_n(\tau) \sim e^{-2\pi y}, \quad y \to \infty.
\end{equation}
The continuous spectrum of the Laplacian can similarly be expanded in an orthonormal basis consisting of the \textbf{real analytic Eisenstein series} with $s$, which labels their eigenvalue, on the critical line $\re s = \half$,
\begin{equation}
\{E_{s=\half+i\RR}\} :\, \Delta_\tau E_s(\tau) = s(1-s) E_s(\tau).
\end{equation}
Unlike the cusp forms, the Eisenstein series do not decay exponentially at the cusp:
\begin{equation}
	E_s(\tau) \sim y^s + {\Lambda(1-s)\over \Lambda(s)}y^{1-s},\quad y\to\infty
\end{equation}
where $\Lambda(s)$ is a symmetrized version of the Riemann zeta function that satisfies a functional equation
\begin{equation}\label{eq:LambdaDefinition}
\begin{aligned}
	\Lambda(s) &\coloneqq \pi^{-s}\Gamma(s)\zeta(2s)\\
	&= \Lambda\(\half-s\).
	\end{aligned}
\end{equation}
We summarize some basic facts about real analytic Eisenstein series, Maass cusp forms and their Fourier decompositions in Appendix \ref{app:eisensteinCusp}.

Any square-integrable modular-invariant function $f(\tau)\in L^2(\mathcal{F})$ admits the following \textbf{Roelcke-Selberg spectral decomposition} into these eigenfunctions \cite{Terras_2013}:
\begin{equation}\label{eq:RoelckeSelberg}
	f(\tau) = \sum_{n=0}^\infty \frac{(f,\nu_n)}{(\nu_n, \nu_n)}\nu_n(\tau) + {1\over 4\pi i}\int_{\re s = \half}ds\, (f,E_s)E_s(\tau).
\end{equation} 

\subsection{The Rankin-Selberg transform}\label{subSec:RankinSelberg}

In this paper the inner product of various partition functions $Z$ with the real analytic Eisenstein series, $(Z,E_s)$, will play a central role. We will see that this quantity amounts to a Mellin transform of the scalar component of the partition function, and inherits many interesting analytic properties in the $s$ plane from those of the Eisenstein series. These observations date back to work of Rankin \cite{rankin_1939} and Selberg \cite{selberg1940bemerkungen}, whose work was usefully extended by Zagier \cite{zbMATH03796039}.\foot{See e.g. \cite{Pioline:2014bra,Green:2014yxa,DHoker:2019mib,DHoker:2019txf} for some more recent applications of this method in the string theory context.}

Consider a modular-invariant function $f(\tau)$ that is of ``rapid decay'', meaning that $f$ decays faster than any polynomial at the cusp $y=\infty$. We will define the \textbf{Rankin-Selberg (RS) transform} of $f$ by its integral over the fundamental domain, weighted by a real analytic Eisenstein series
\begin{equation}\label{eq:RankinSelberg}
\begin{aligned}
	R_s[f] &\coloneqq \int_{\mathcal{F}}{dxdy\over y^2}\,f(\tau) E_s(\tau)\\
	&= \int_{\mathcal{F}}{dx dy\over y^2}\, f(\tau) \sum_{\gamma\in\Gamma_\infty\backslash PSL(2,\ZZ)}\im(\gamma\tau)^s\\
	&= \int_{\Gamma_\infty\backslash \HH}{dxdy\over y^2}\, f(\tau) y^s\\
	&= \int_0^\infty dy \, y^{s-2}f_0(y)
\end{aligned}
\end{equation}
where $f_0(y)$ is the zeroth Fourier component of $f$
\begin{equation}
	f_0(y) = \int_{-\half}^{\half}dx\, f(\tau).
\end{equation}
In (\ref{eq:RankinSelberg}) we have made use of a standard \emph{unfolding trick}, where we employ the fact that the Eisenstein series is a Poincar\'e series. The Mellin transform in the last line of (\ref{eq:RankinSelberg}) converges provided $\re s > 1$. However as described in Appendix \ref{subApp:eisenstein}, the Eisenstein series has a meromorphic continuation in $s$ that is holomorphic everywhere in the $s$ half-plane to the right of the critical line $\re s = \half$ except for a simple pole with constant residue ${3\over \pi}$ at $s=1$, thus endowing the RS transform with its own meromorphic continuation. The Eisenstein series also satisfies the functional equation (\ref{eq:eisensteinFunctionalEqn}), which implies a functional equation for the RS transform
\begin{equation}
	\Lambda(s)R_s[f] = \Lambda(1-s)R_{1-s}[f].	
\end{equation}
Moreover, the residue of the RS transform at the pole $s=1$ encodes the average of the function $f$ over the fundamental domain
\begin{equation}\label{avgrs}
	\Res_{s=1}R_s[f] = {3\over \pi}\int_{\mathcal{F}}{dxdy\over y^2}\, f(\tau).
\end{equation}

However, we will see that CFT primary partition functions are not of rapid decay, and are thus not subject to the spectral analysis as stated in (\ref{eq:RoelckeSelberg}). Thankfully, Zagier has developed the RS method to accommodate modular functions that are of ``slow growth'' at the cusp \cite{zbMATH03796039}. We will summarize many of the relevant details from \cite{zbMATH03796039} here. When we say that a modular-invariant function $f$ is of slow growth at the cusp, what we mean is 
\begin{equation}
	f(\tau) \sim \varphi(y) + \text{(sub-polynomial)}
	,\quad y\to\infty,
\end{equation}
where $\varphi(y)$ grows sub-exponentially,
\begin{equation}\label{eq:varphiDefinition}
	\varphi(y) = \sum_{i=1}^m {c_i\over n_i!}y^{\alpha_i}\log^{n_i}y\quad (c_i, \alpha_i \in \mathbb{C}\,,~n_i \in \mathbb{Z}_{\geq 0}).
\end{equation}
In Section \ref{sec:narain}, we will see that the primary partition functions of Narain lattice CFTs correspond to the maximally simple $m=1,\, c_1=1,\, n_1 = 0,\, \alpha_1 = {c\over 2}$. Then the definition of the RS transform is simply modified to omit the problematic terms at the cusp, namely
\begin{equation}\label{RSsub}
	R_s[f] \coloneqq \int_0^\infty dy\, y^{s-2}\left(f_0(y)-\varphi(y)\right).
\end{equation}
This converges when $\re s$ is sufficiently large. This procedure, which essentially amounts to throwing out the problematic terms in the scalar part of $f$, may seem ad hoc, but it turns out to be quite natural. The reason is that this definition of the RS transform can be shown \cite{zbMATH03796039} to be equivalent to a renormalized integral of the product of the function weighted by the Eisenstein series over the fundamental domain
\begin{equation}
\begin{aligned}
	R_s[f] &= \int_{\mathcal{F}_T} {dxdy\over y^2} f(\tau)E_s(\tau) + \int_{\mathcal{F}-\mathcal{F}_T} {dxdy\over y^2}\left(f(\tau)E_s(\tau) - \varphi(y)\varphi_s(y)\right) - h^{(T)}_s - {\Lambda(1-s)\over \Lambda(s)}h^{(T)}_{1-s}\\
	&\coloneqq \rnrn\left(\int_{\mathcal{F}}{dxdy\over y^2}f(\tau)E_s(\tau)\right)
	\label{eq:ZRS}
\end{aligned}
\end{equation}
where $\mathcal{F}_T = \mathcal{F}\cap \{y\le T\}$ is the fundamental domain with an infrared cutoff, $\varphi_s(y)$ is the zeroth Fourier mode of the Eisenstein series
\begin{equation}
	\varphi_s(y) \coloneqq y^s + {\Lambda(1-s)\over\Lambda(s)}y^{1-s} 
\end{equation}
and 
\begin{equation}
	h^{(T)}_s \coloneqq \int_0^T dy\, y^{s-2}\varphi(y).
\end{equation}

This modification of the RS transform will be both conceptually and technically useful in our spectral analysis of CFT partition functions. To explain why, consider the case of the constant function, $f(\tau) = 1$. In this case $f_0(y) = \varphi(y) = 1$ so its RS transform trivially vanishes. In particular one has
\begin{equation}
	R_s[1] = \int_{\mathcal{F}}{dx dy\over y^2}E_s(\tau) = 0,\quad 0<\re s < 1.
\end{equation}
Similarly, if we consider the case that $f$ is an Eisenstein series, $f(\tau) = E_{r}(\tau)$, then again its RS transform \eqr{RSsub} vanishes because $f_0(y) = \varphi(y) = \varphi_r(y)$:
\begin{equation}
	R_s[E_r(\tau)] = \rnrn\left(\int_{\mathcal{F}}{dxdy\over y^2}E_r(\tau)E_s(\tau)\right) = 0.
\end{equation}
This motivates the following interpretation of the renormalized integral of $f$ in the fundamental domain. Since $R_s[E_r(\tau)]=0$, we can subtract from $f(\tau)$ a suitable combination of Eisenstein series without affecting its renormalized integral,
\begin{equation}\label{eq:fTildeDefinition}
	\widetilde f(\tau) \coloneqq f(\tau) - \sum_{i|\alpha_i\ge \half}\left.c_i{\partial^{n_i}\over\partial s^{n_i}}E_s(\tau)\right|_{s=\alpha_i},
\end{equation}
such that the resulting function is square integrable:
\begin{equation}
	R_s[f] = \int_{\mathcal{F}}{dxdy\over y^2}\, \widetilde{f}(\tau) E_s(\tau).
\end{equation}
This converges for $\re s$ in a certain range dictated by the powers of $y$ in $\varphi(y)$. The RS transform again inherits the meromorphic continuation of the Eisenstein series in the $s$ plane, now with poles in the plane to the right of $\re s = \half$ at $s=1$ and $s=\max(\alpha_i,1-\alpha_i)$. 

In other words -- and as groundwork for the viewpoint espoused later -- one simply forms a new, square-integrable modular-invariant function $\widetilde f (\t)$ by subtracting the appropriate linear combination of $E_s(\t)$ and its derivatives from $f(\t)$, one for each term in \eqr{eq:varphiDefinition}.\foot{The reader may notice that terms with $\alpha_i < 1/2$ are not subtracted off in \eqr{eq:fTildeDefinition}, but are nevertheless included in the definition of $\varphi(\t)$. They are so included because they spoil convergence of the original RS transform for some range of $s$. However, it also follows from Zagier's theorem that for RS transforms with $\re s=1/2$ -- the only value needed for spectral decomposition -- the terms with $\alpha_i < 1/2$ may be safely neglected in the RS transform. This is consistent, as it must be, with the fact that they do not violate square-integrability, so $\widetilde f(\t)$ must admit a Roelcke-Selberg spectral decomposition without any modifications.} The procedure of mapping the divergent terms to modular-invariant functions with the same asymptotic growth is what we will call ``modular completion.'' The new function $\widetilde f(\t)$ then admits a spectral decomposition \eqr{eq:RoelckeSelberg}, with the overlap integrals computed using the unmodified RS transform. Thus, the form of the spectral decomposition for functions of slow growth is
\begin{equation}\label{eq:modifiedRoelckeSelberg}
	f(\tau) = \sum_{i|\alpha_i\ge \half}\left.c_i{\partial^{n_i}\over\partial s^{n_i}}E_s(\tau)\right|_{s=\alpha_i} + \sum_{n=0}^\infty \frac{( f,\nu_n)}{(\nu_n, \nu_n)}\nu_n(\tau) + {1\over 4\pi i}\int_{\re s = \half}ds\, R_{1-s}[\widetilde f]E_s(\tau).
\end{equation}
 which is equivalent to (\ref{eq:RoelckeSelberg}) applied to the function $\widetilde f(\t)$. 

\subsection{Comparison with harmonic analysis on the Euclidean conformal group}\label{eucharm}

Harmonic analysis on the Euclidean conformal group $SO(d+1,1)$ \cite{Dobrev:1977qv} is a powerful tool for the study of the constraints of conformal invariance on four-point functions of local operators, that has underlied many recent developments in the conformal bootstrap \cite{Cornalba:2006xm,Costa:2012cb,Cornalba:2007fs,Murugan:2017eto,Gadde:2017sjg,Hogervorst:2017sfd,Caron-Huot:2017vep,Hogervorst:2017kbj,Kravchuk:2018htv,Karateev:2018oml,Liu:2018jhs}, and has some parallels with the technology developed in this section. Here we briefly review some elements of $SO(d+1,1)$ harmonic analysis in order to contextualize our results.

In Euclidean kinematics, the four-point function of local operators $\langle \phi_1\phi_2\phi_3\phi_4\rangle$, taken to be scalars for simplicity, is, up to a kinematic prefactor, a function $G(z,\bar z)$ of a single complex conformally-invariant cross-ratio $z$. The function $G(z,\bar z)$ admits a conformal block decomposition
\begin{equation}
	G(z,\bar z) = \sum_{\Delta,\, j} C_{12\mathcal{O}}C_{34\mathcal{O}}F_{\Delta, j}(z,\bar z),
\end{equation}
where the sum is over local primaries $\mathcal{O}$ with dimension $\Delta$ and spin $j$ appearing in the $\phi \times \phi$ OPE, $F_{\Delta, j}$ are the conformal blocks that encode the kinematical contribution of $\mathcal{O}$ and its descendants, and the coefficients $C$ are the corresponding structure constants. Harmonic analysis on the conformal group allows one to decompose the function $G(z,\bar z)$ into a complete basis of eigenfunctions of the conformal Casimir that are single-valued functions of the cross-ratio in Euclidean kinematics, namely, the conformal partial waves $\Psi_{\Delta, j}$ (which are a particular linear combination of the conformal block $F_{\Delta, j}$ and the ``shadow block'' $F_{d-\Delta,j}$) \cite{Dobrev:1977qv,Costa:2012cb,Caron-Huot:2017vep,Simmons-Duffin:2017nub,Hogervorst:2017sfd}. For real external dimensions, 
\begin{equation}\label{eq:CPWDecomposition}
	G(z,\bar z) = \text{(non-normalizable)} + \sum_{j=0}^\infty\,\int_{{d\over 2}}^{{d\over 2}+i\infty}{d\Delta\over 2\pi i}\, C_j(\Delta)\Psi_{\Delta, j}(z,\bar z).
\end{equation}
In the above equation we have emphasized that the contributions of certain operators in the OPE, in particular those with $\Delta \le {d\over 2}$ (which always includes the identity operator), must be subtracted from the four-point function in order to apply the spectral decomposition and preserve normalizability. The coefficient function $C_j(\Delta)$ is a sort of resolvent for the structure constants, in the sense that it has simple poles in $\Delta$ corresponding to the dimensions of operators in the OPE\footnote{Below we have assumed a natural normalization of the conformal partial waves.}
\begin{equation}\label{eq:coefficientFunctionResidue}
	-\Res_{\Delta = \Delta_i} C_{j_i}(\Delta) = C_{12\mathcal{O}_i}C_{34\mathcal{O}_i},\quad \text{for generic $\Delta_i$}.
\end{equation}
We say that this applies for ``generic'' $\Delta_i$ due to the fact that the conformal partial waves have poles in the right half $\Delta$ plane with residues that are themselves conformal blocks with special values of the dimension (which may thus shift the relationship between the structure constants and the residue of the coefficient function), and due to the need for subtractions of states that give non-normalizable contributions. That the conformal partial waves are orthogonal with respect to a certain inner product means that one can extract the coefficient function $C_j(\Delta)$ by means of a \emph{Euclidean inversion formula} \cite{Caron-Huot:2017vep}
\begin{equation}\label{eq:euclideanInversion}
	C_j(\Delta) = N_j(\Delta) \int_{\mathbb{C}}d^2 z \, \mu(z,\bar z) \Psi_{\Delta, j}(z,\bar z) G(z,\bar z),
\end{equation}
where $N_j(\Delta)$ is a proportionality constant and $\mu(z,\bar z)$ a measure, whose detailed form are known but unimportant for our purposes. 

Let us now make the analogy with harmonic analysis on the fundamental domain. One may think of the modified Roelcke-Selberg decomposition (\ref{eq:modifiedRoelckeSelberg}) as analogous to (\ref{eq:CPWDecomposition}). As we will see when studying the spectral decomposition of both the Narain theories, as discussed in Section \ref{sec:narain}, and especially in more general CFTs in Section \ref{sec4}, we will also encounter the need to subtract non-normalizable contributions: in particular, those corresponding to the {\it light} primary operators. The RS transform (\ref{eq:RankinSelberg}) discussed in the previous subsections, which extracts the overlap coefficient $(f,E_s) = R_{1-s}[f]$ that appears in the Roelcke-Selberg decomposition (\ref{eq:RoelckeSelberg}) from an integral of $f$ weighted by an Eisenstein series over the fundamental domain, is closely analogous to this Euclidean inversion formula, but with some key differences.

For one, an obvious difference from (\ref{eq:CPWDecomposition}) is that in applying harmonic analysis on the fundamental domain, we are expanding the partition function into a complete basis of functions that are themselves modular-invariant, unlike the conformal partial waves that appear above which are not crossing-symmetric.\foot{A crossing-symmetric decomposition of CFT correlators was proposed in \cite{gopak} (see also \cite{Mazac:2018qmi,Maloney:2016kee}). It would be productive to further explore the analogy between these proposals and the present work.} This is related to the fact that a key feature of the Euclidean inversion formula for correlators is that the eigenfunctions are labeled by their spin $j$, which is restricted to be an integer.  By contrast, the eigenfunctions in the Roelcke-Selberg decomposition do not have such a spin label, but rather are all certain infinite sums over spins.  In fact, once one has determined the Roelcke-Selberg decomposition, one can use it to evaluate the partition function for complex values of $x$ and $y$ as long as $\textrm{Re}(y)>0$, so in this sense the decomposition is equally Lorentzian or Euclidean, even though our method for extracting it relies on the partition function in the Euclidean regime $(x,y) \in \mathbb{H}$.  Another important difference stemming from the built-in modular invariance is that there is no immediate analog of (\ref{eq:coefficientFunctionResidue}) that directly relates the RS transform of a partition function to the degeneracies of particular operators in the spectrum. However, after some subtractions or if $Z$ has particularly tame asymptotic growth, we are able to extract a resolvent from $R_s[Z]$ via a formal Laplace transform (e.g. for the $c=1$ free boson; see Appendix \ref{app:c=1Resolvent}).  A final difference is the presence of the infinite family of cusp forms in the Roelcke-Selberg decomposition, which do not have any obvious analog in the decomposition into conformal partial waves.

In \cite{Caron-Huot:2017vep}, by exploiting an analogy with the Froissart-Gribov formula for the partial wave coefficients in scattering amplitudes, Caron-Huot was able to deform the integration region in the Euclidean inversion formula (\ref{eq:euclideanInversion}) to one over the Lorentzian causal diamond $z,\bar z \in (0,1)$ with profound implications for the structure of the spectrum of a CFT. In particular, by providing a formula for the CFT data (for operators of sufficiently large spin) that is analytic in the spin $j$, the result of \cite{Caron-Huot:2017vep} formalized the idea that the local operator content of a CFT is organized into Regge trajectories of increasing spin. This implies a remarkable rigidity of the spectrum. Although our methods for $SL(2,\ZZ)$ are apparently Euclidean, we will nevertheless see in Section \ref{secspecdet} that the spectral analysis of 2d CFT partition functions has conceptually similar implications for the rigidity of the spectrum of local operators.

\section{Application to Narain lattice CFTs}\label{sec:narain}
In this section we will discuss the application of techniques from harmonic analysis on $\mathcal{F}$ to the study of partition functions of free boson CFTs based on an even self-dual lattice $\Lambda\in \RR^{c,c}$. Narain lattice CFTs are characterized by a $U(1)^c\times U(1)^c$ current algebra, and thus provide a natural testing ground for the application of the technology described in the previous section. The reason is the following. For unitary CFTs with two copies of a $U(1)^c$ chiral algebra, the primary partition function $Z_p(\tau)$ that counts $U(1)^c\times U(1)^c$ primary operators is given in terms of the full partition function $Z(\tau)$ by
\begin{equation}
	Z_p(\tau) \coloneqq \left(y^{1/2}|\eta(\tau)|^2\right)^c Z(\tau),
	\label{eq:stripoffcrap}
\end{equation}
which only diverges polynomially at the cusp
\begin{equation}
	Z_p(\tau) \sim y^{c\over 2} + \mathcal{O}(y^{c\over 2}e^{-2\pi \Delta_{*} y}), \quad y \to \infty
\end{equation}
where $\Delta_{*}$ is the dimension of the lightest non-trivial primary operator in the spectrum. The divergence is merely polynomial, rather than exponential, because Narain CFTs have central charge equal to the number of currents; more precisely, $c=c_{\rm currents}$, where $c_{\rm currents}$ parameterizes the asymptotic density of current composites as defined in \eqr{ccurr}. Moreover, the only source of this divergence is the universal contribution of the identity operator. Thus the primary partition function is a modular-invariant function of ``slow growth'' on the fundamental domain, so we can apply the technology of Section \ref{subSec:RankinSelberg} to understand its spectral decomposition. In contrast, partition functions of generic CFTs diverge {\it exponentially} at the cusp, and so in that case more care must be taken to make sense of a spectral decomposition, as we discuss further in Section \ref{sec4}.

In the remainder of this section we drop the subscript on the primary partition function $Z_p(\tau)$, leaving the restriction to primaries implicit.  

\subsection{Spectral decomposition of specific partition functions}
We will now explicitly compute the spectral decomposition of Narain CFT partition functions, $Z^{(c)}$, at small values of $c$, and will outline the general computation for general $c$. We do so by explicit computation of the RS transform $(Z^{(c)},E_s) = R_{1-s}[Z^{(c)}]$ -- which, as we have seen in Section \ref{subSec:RankinSelberg}, amounts to a Mellin transform of the scalar sector of the partition function -- and also computation of $(Z^{(c)},\nu_n)$, the inner product with the cusp forms.

Narain lattices are far from unique; a particular lattice is specified by a point in Narain's moduli space $\mathcal{M}_c$
\begin{equation}
	\mathcal{M}_c = O(c,c;\ZZ)\backslash O(c,c;\RR) / O(c)\times O(c).
\end{equation}
Such theories are realized as sigma models with toroidal target space $T^c$, with the data of the Narain lattice encoded in the $c^2$ degrees of freedom of the metric and $B$-field flux of the target. The modular-invariant primary partition function for a Narain lattice CFT with central charge $c$ is given by the following sum over momentum and winding modes
\begin{equation}\label{eq:narainPartitionFunction}
	Z^{(c)}(\tau;m) = y^{c\over 2} \sum_{n_a,w^a\in\ZZ^c}\exp\left(-\pi y M_{n,w}(m)^2 + 2\pi i x n\cdot w\right),
\end{equation}
where $m$ denote the target space moduli (the metric $G$ and 2-form flux $B$) and\footnote{Throughout this paper we set $\alpha'=1$. At $c=1$ for instance we take $r=1$ to be the self-dual radius.}
\begin{equation}
	M_{n,w}(m)^2 \coloneqq G^{ab}\left(n_a+B_{ac}w^c\right)\left(n_b+B_{bd}w^d\right)+G_{cd}w^cw^d.
\end{equation}

From their explicit forms, one can confirm that the Narain partition functions satisfy the differential equation \cite{Obers:1999um,Maloney:2020nni}
\begin{equation}
	\left(\Delta_\tau - \Delta_{\mathcal{M}_c} -{c\over 2}\left(1-{c\over 2}\right)\right)Z^{(c)}(\tau;m) = 0,
	\label{eq:DIFFEQ}
\end{equation}
where
\begin{equation}
	\Delta_{\mathcal{M}_c} = -G_{ac}G_{bd}\left(\widehat\partial_{G_{ab}}\widehat\partial_{G_{cd}} + {1\over 4}\partial_{B_{ab}}\partial_{B_{cd}}\right) - G_{ab}\widehat\partial_{G_{ab}}
\end{equation}
is the Laplacian associated with the Zamolodchikov metric on the target space moduli space, with $\widehat \partial_{G_{ab}} \coloneqq \half\left(1+\delta_{ab}\right)\partial_{G_{ab}}$.

\subsubsection{\texorpdfstring{$c=1$}{c1}}

Consider the special case of the $c=1$ free boson compactified on a circle of radius $r$. In this case the primary partition function is given by
\begin{equation}\label{eq:c=1}
	Z^{(c=1)}(\tau;r) = \sqrt{y}\sum_{n,w\in\ZZ}\exp\left(-\pi y\left(n^2 r^{-2}+w^2 r^2\right)+2\pi i x nw\right).
\end{equation}
We begin by computing the continuous part of the spectral decomposition via RS transform of the partition function (\ref{eq:c=1}), using technology of Section \ref{subSec:RankinSelberg} for modular functions of ``slow growth,'' followed by the discrete part. 

Neglecting the contribution of the identity operator by defining
\begin{equation}
	\widetilde Z^{(c=1)}(\tau;r) \coloneqq \sqrt{y}\sideset{}{'}\sum_{n,w\in\ZZ}\exp\left(-\pi y\left(n^2 r^{-2}+w^2 r^2\right)+2\pi i x nw\right),
\end{equation} 
where the prime on the summation denotes the omission of $n=w=0$, we get
\begin{equation}
\begin{aligned}
	R_s[Z^{(c=1)}] &= \int_{\mathcal{F}}{dxdy\over y^2}\widetilde Z^{(c=1)}(\tau;r)E_s(\tau) \\
	&= 2\int_0^\infty dy \, y^{s-{3\over 2}}\sum_{n=1}^\infty\left(e^{-\pi y n^2 r^2}+e^{-\pi y n^2 r^{-2}}\right).
\end{aligned}
\end{equation}
Provided the real part of $s$ is sufficiently large (in particular, $\re s > \half$), we can exchange the sum and the integral, leading to
\begin{equation}
\begin{aligned}\label{eq:c=1RS}
	R_s[Z^{(c=1)}] &= 2\sum_{n=1}^\infty (\pi n^2)^{\half-s}\Gamma\left(s-\half\right)\left(r^{1-2s}+r^{2s-1}\right)\\
	&= 2\Lambda\left(s-\half\right)\left(r^{1-2s}+r^{2s-1}\right).
\end{aligned}
\end{equation}
The result indeed has poles at $s=1$ and $s=\half$ as promised. We thus conclude that
\begin{equation}\label{eq:c=1EisensteinOverlap}
	(Z^{(c=1)},E_s) = R_{1-s}[Z^{(c=1)}] =  2\Lambda(s)\left(r^{1-2s}+r^{2s-1}\right).
\end{equation}

Next, the constant part of the spectral decomposition may be computed by applying \eqr{avgrs}:
\begin{equation}\label{eq:c=1Constant}
	\Res_{s=1}\left(R_s[Z^{(c=1)}]\right) = r + r^{-1}.
\end{equation}
In fact, the results (\ref{eq:c=1RS}) and (\ref{eq:c=1Constant}) have previously appeared in the literature in \cite{Angelantonj:2011br}, where the RS method was used as a technical tool for the computation of certain one-loop integrals in toroidal compactifications of string theory. For those purposes, only the residue of the RS transform at $s=1$ was desired.

So far we have computed the constant and continuous components in the spectral decomposition of the primary partition function of the $c=1$ free boson. To finish the job we need to add in both the contribution of the vacuum state and the Maass cusp forms. In fact it turns out that both of these contributions will vanish. For the former, we see in the modified Roelcke-Selberg decomposition (\ref{eq:modifiedRoelckeSelberg}) with $\alpha_i=1/2$ and $n_i=0$ leaves $E_{\half}(\tau)$, which is in fact zero: $E_{\half}(\tau)= 0$. For the latter, in Appendix 
\ref{app:derivingEight} we explicitly show that the Maass cusp forms do not show up at $c=1$ by computing the overlap $(Z^{(c=1)}, \nu_n)$ and showing it vanishes. Moreover, we have verified this by checking numerically that $(Z^{(c=1)},\nu_n)$ can be made seemingly arbitrarily small by increasing the precision of the numerical integration. Thus, the spectral representation of the partition function of the $c=1$ free boson is given exactly by 
\begin{equation}\label{eq:c=1Spectral}
	\boxed{Z^{(c=1)}(\tau;r) = r + r^{-1} + {1\over 4\pi i}\int_{\re s = \half}ds\, 2\Lambda(s)\left(r^{1-2s}+r^{2s-1}\right)E_s(\tau).}
\end{equation}
See Appendix \ref{app:c=1Resolvent} to see explicitly that the resolvent inherited from (\ref{eq:c=1Spectral}) has poles precisely at the locations of physical operators in the free boson CFT with residues equal to the correct degeneracies. 
 
The partition functions of the $\ZZ_2$ orbifolds of the compact boson admit a similarly straightforward spectral decomposition. The orbifold partition functions, $Z^{(c=1)}_{\rm orb}(\tau;r)$, can be written in terms of free boson partition functions as
\begin{equation}\label{ZZ2}
	Z^{(c=1)}_{\rm orb}(\tau;r) = \half Z^{(c=1)}(\tau;r) + Z_{\rm tw}(\tau),
\end{equation}
where the contribution of the $\ZZ_2$ twisted sectors is independent of the target space moduli:
\begin{equation}
	Z_{\rm tw}(\tau) = \half \sqrt{y}\left(|\theta_2\theta_3|+|\theta_3\theta_4|+|\theta_4\theta_2|\right),
\end{equation}
where $\theta_i$ are elliptic theta functions. Then the well-known relation 
\be
Z^{(c=1)}_{\rm orb}(\tau;r=1) = Z^{(c=1)}(\tau;r=2) 
\ee
implies
\begin{equation}\label{twspec}
	Z_{\rm tw}(\tau) = {3\over 2} + {1\over 4\pi i}\int_{\re s = \half}ds\, 2\Lambda(s)\left(2^{1-2s}+2^{2s-1} - 1\right)E_s(\tau)
\end{equation}
The spectral decomposition of $Z^{(c=1)}_{\rm orb}(\tau;r)$ then follows from \eqr{eq:c=1Spectral}, \eqr{ZZ2} and \eqr{twspec}.\footnote{Of course, since we are computing $y^{1/2} |\eta(\tau)|^2$ times the orbifold partition function, and $\frac1{\eta(\tau)}$ is not the vacuum character for the $S^1/\mathbb{Z}_2$ theory, the spectral decomposition above, unlike (\ref{eq:c=1Spectral}), does not have a positive inverse Laplace transform (although it is still discrete).}

\subsubsection{\texorpdfstring{$c=2$}{c2}}

We next consider the family of $c=2$ Narain CFTs with $T^2$ target space. The moduli space of such theories is captured by two elements $\sigma, \rho$ of $\HH/PSL(2,\ZZ)$, respectively the complex structure and complexified K\"ahler structure of the target. They are given in terms of the metric and $B$ field flux on the $T^2$ as (see e.g. \cite{Dijkgraaf:1987jt}) 
\begin{equation}
\begin{aligned}\label{eq:T2Moduli}
	\rho &= B + i {\sqrt{\det G}} \\
	\sigma &= {G_{12}\over G_{11}} + i {\sqrt{\det G}\over G_{11}}.
\end{aligned}
\end{equation}
In these variables, the primary partition function can be written as\footnote{This expression can be elegantly rewritten in terms of Poincar\'e series and Hecke operators, see Appendix \ref{app:derivingEight}.} 
\begin{equation}
\begin{aligned}\label{eq:c=2Narain}
	Z^{(c=2)}(\tau;\rho,\sigma) = y \sum_{n,w\in\mathbb{Z}^2} \exp\Bigg[{\pi i \over 2}\Bigg(&{\tau\over \rho_2\sigma_2}\left|n_2-n_1 \sigma -\rho(w^1+w^2 \sigma)\right|^2 \\- &{\bar\tau\over \rho_2\sigma_2}\left|n_2-n_1\bar\sigma - \rho\left(w^1+w^2\bar\sigma\right) \right|^2 \Bigg)\Bigg].
\end{aligned}
\end{equation}
Remarkably, $Z^{(c=2)}(\tau; \rho, \sigma)$ obeys a ``triality" symmetry: it is totally symmetric under exchange of the three moduli \cite{Dijkgraaf:1987jt}. The invariance under exchange of $\rho$ and $\sigma$ is a special case of mirror symmetry.  The invariance under the exchange of the worldsheet modular parameter $\tau$ with the target space moduli $\rho,\sigma$, which follows from Poisson summation, is more mysterious.  

The RS transform of the $c=2$ Narain partition function is given by
\begin{equation}
\begin{aligned}
	R_s[Z^{(c=2)}] &= \int_{\mathcal{F}}{dxdy\over y^2}\, \widetilde Z^{(c=2)}(\tau;\rho,\sigma) E_s(\tau) \\
	&= \int_0^\infty dy\, y^{s-1}\sideset{}{'}\sum_{n,w\in\ZZ^2}\delta_{n\cdot w,0}\exp\left(-\pi y M_{n,w}(m)^2\right)\\
	&= \pi^{-s}\Gamma(s) \sideset{}{'}\sum_{n,w\in\ZZ^2}{\delta_{n\cdot w,0}\over\left(M_{n,w}(m)^2\right)^{s}},
\end{aligned}
\end{equation}
where again the prime on the summation indicates the omission of the identity operator $n_a = w^a = 0$ and we have assumed $\re s > 0$ in order to exchange the integral and the sum. 

In fact, this constrained sum was worked out very explicitly in \cite{Angelantonj:2011br}, whose results we briefly summarize. In solving the constraint $n\cdot w = 0$, there are essentially two cases to consider. In the first case we simply set $w^a = 0$, and we have
\begin{equation}
	M_{(n_1,n_2),(0,0)}(m)^2 = {|n_2-\sigma n_1|^2\over \sigma_2\rho_2}.
\end{equation}
In the second case we set
\begin{equation}
	(n_1,n_2,w^1,w^2) = (du_1,du_2, -cu_2, cu_1),
\end{equation}
where $u_1,u_2\in\ZZ$ with $(u_1,u_2) = 1$ and $c,d\in\ZZ$ with $c\ge 1$. In this case we have
\begin{equation}
	M_{(du_1,du_2),(-cu_2,cu_1)}(m)^2 = {|u_2-\sigma u_1|^2 |c\rho+d|^2\over \sigma_2\rho_2}.
\end{equation} 
The RS transform then takes the form
\begin{equation}\label{eq:c=2RS}
\begin{aligned}
	R_s[Z^{(c=2)}] &= \pi^{-s}\Gamma(s)\left[\sideset{}{'}\sum_{n_1,n_2\in\mathbb{Z}}\left({\rho_2\sigma_2\over |n_2-\sigma n_1|^2}\right)^s + \sideset{}{'}\sum_{\substack{u_1,u_2\in\ZZ \\ (u_1,u_2)=1}}\left(\sigma_2\over |u_2-\sigma u_1|^2\right)^s\sum_{\substack{c,d\in\ZZ \\ c \ge 1}}\left({\rho_2\over |c\rho+d|^2}\right)^s\right]\\
	&= 2\Lambda(s)E_s(\rho)E_s(\sigma).
\end{aligned}
\end{equation}
So the inner product of the $c=2$ Narain partition functions with the real analytic Eisenstein series is given by the following product of Eisenstein series
\begin{equation}\label{eq:c=2InnerProduct}
\begin{aligned}
	(Z^{(c=2)},E_s) = R_{1-s}[Z^{(c=2)}] &= 2\Lambda(1-s)E_{1-s}(\rho)E_{1-s}(\sigma) \\&= 2{\Lambda(s)^2\over \Lambda(1-s)}E_s(\rho)E_s(\sigma).
\end{aligned}
\end{equation}

At the cusp, the $c=2$ Narain partition functions diverge linearly, $Z^{(c=2)}(\tau;m) \stackrel{y\to\infty}{\sim} y$. And indeed we see as expected that the RS transform (\ref{eq:c=2RS}) has a double pole at $s=1$. For this reason we must take slightly more care in extracting the constant part of the spectral decomposition. Because of the linear divergence of $Z^{(c=2)}$ at the cusp, in order to apply the Roelcke-Selberg theorem we subtract $\widehat E_1(\tau) - \omega$ from the partition function, where $\widehat E_1(\t)$ is the non-singular part of the Eisenstein series at $s=1$ (see (\ref{eq:e1hatdef}))
\be
\widehat{E}_{1}(\tau) = \lim_{s\rightarrow1}\left[ E_s(\tau) -\frac{3}{\pi(s-1)}\right]
\ee  
and $\omega \coloneqq {6\over \pi}\left(1-12\zeta'(-1)-\log(4\pi)\right)$ is the constant part of $\widehat E_1(\t)$. The remainder of the constant part of the spectral decomposition comes from the residue of the RS transform at $s=1$, cf. \eqr{avgrs}, which is given by
\begin{equation}
	\Res_{s=1}\left(R_s[Z^{(c=2)}]\right) = \widehat{E}_1(\rho) + \widehat{E}_1(\sigma) + \delta,
\end{equation}
where $\delta \coloneqq {18\o \pi^2} \Lambda'(1) = {3\over \pi}\left(\gamma_E+\log(4\pi) + 24 \zeta'(-1) -2\right)$ with $\gamma_E$ the Euler-Mascheroni constant. 

Finally we address the discrete part of the spectrum. Following the discussion in Appendix \ref{app:eisensteinCG}, we generally expect all $c>1$ Narain partition functions to have support on the cusp forms. The differential equation (\ref{eq:DIFFEQ}), the triality of the $c=2$ partition function, and the orthogonality of the cusp forms together imply that the discrete contribution must take the following highly constrained form:
\begin{align}
	& Z^{(c=2)}(\tau; \rho, \sigma)  \supset \sum_{n=1}^\infty c^+_n (\nu^+_n,\nu^+_n)^{-1}\nu^+_n(\rho)\nu^+_n(\sigma)\nu^+_n(\tau) +  \sum_{n=1}^\infty c^-_n (\nu^-_n,\nu^-_n)^{-1}\nu^-_n(\rho)\nu^-_n(\sigma)\nu^-_n(\tau), \label{eq:blahblahblahtraility} 
\end{align}
where $c_n^\pm$ are a set of constants and where we have separated the cusp forms into the even and odd cusp forms, denoted by $\nu^+$ and $\nu^-$ respectively.\footnote{A previous version of this paper incorrectly omitted the odd cusp forms $\nu_n^-$ from the spectral decomposition. However, Narain theories with $B$ field turned on are generically not parity invariant, which imply the presence of odd cusp forms. We are grateful to Ying-Hsuan Lin and Yifan Wang for discussions related to this point.} Any other product of cusp forms totally symmetric in $\rho, \sigma, \tau$ will not obey (\ref{eq:DIFFEQ}). We have also normalized the $c_n^\pm$'s in (\ref{eq:blahblahblahtraility}) by a factor of $(\nu^\pm_n, \nu^\pm_n)$ which will turn out to be convenient.

In Appendix \ref{app:derivingEight} we analytically derive:\footnote{One may be concerned that the overlap coefficient with the odd Maass cusp forms is imaginary, given that the cusp forms are real functions when $\overline\tau = \tau^*$. This is not a problem, since parity-non-invariant partition functions need not be real. Indeed, we see below that the overlap with the odd cusp forms is only non-vanishing when the $B$-field is turned on, in particular when the $c=2$ partition function is not parity invariant.}
\begin{align}
    \left(Z^{(c=2)},\nu^+_n\right) &= 8 \nu^+_n(\rho)\nu^+_n(\sigma) \nonumber \\
        \left(Z^{(c=2)},\nu^-_n\right) &= -8i \nu^-_n(\rho)\nu^-_n(\sigma) \label{eq:eight}
\end{align}
 by making use of the fact that the $c=2$ Narain partition function (\ref{eq:c=2Narain}) can be written in terms of Poincar\'e series and Hecke operators.

Assembling the continuous and discrete pieces, we thus obtain the following exact form for the spectral decomposition of the $c=2$ Narain partition functions:
\begin{empheq}[box=\fbox]{equation}\label{eq:c=2Spectral}
\begin{split}
	Z^{(c=2)} (\tau;\rho,\sigma) =& \, \alpha + \widehat{E}_1(\rho) + \widehat{E}_1(\sigma) + \widehat{E}_1(\tau)\\
	& \, + {1\over 4\pi i}\int_{\re s =\half}ds\, 2{\Lambda(s)^2\over \Lambda(1-s)}E_s(\rho)E_s(\sigma)E_s(\tau)\\
	& \, + 8\sum_{\epsilon = \pm} \delta_\epsilon \sum_{n=1}^\infty(\nu^\epsilon_n,\nu^\epsilon_n)^{-1}\nu^\epsilon_n(\rho)\nu^\epsilon_n(\sigma)\nu^\epsilon_n(\tau),
\end{split}
\end{empheq}
where the constant $\alpha$ is given by
\begin{align}
	\alpha &\coloneqq \delta - \omega = \frac3\pi \(\gamma_E + 3\log(4\pi) + 48\zeta'(-1) - 4\) 
	\approx -3.6000146,
\end{align}
and $\delta_\epsilon$ is defined as
\begin{align}
\delta_+ \coloneqq 1, ~~~~~\delta_- \coloneqq -i.
\end{align}
The result exhibits manifestly the remarkable triality of \cite{Dijkgraaf:1987jt}. We do not have a physical explanation for the ``8,'' but it is tempting to speculate that it is counting something.

We pause to note that we have also checked our expression (\ref{eq:c=2Spectral}) numerically, in large part thanks to the plethora of numerical data available on the cusp forms.\footnote{We refer the reader to the online database \cite{LMFDB} for numerical data on the cusp forms, and to Appendix \ref{subApp:cusp} for more details of the cusp forms. See also \cite{Booker} for even higher numerical precision on the cusp forms which we used in our fit (\ref{eq:8sgalore}) (we are grateful to A. Sutherland for pointing out this reference) and \cite{Then_2004} for a study of cusp forms at large eigenvalue.} For example, by plugging in various values of $\tau$, $\rho$, and $\sigma$ into $Z^{(c=2)}(\tau; \rho, \sigma)$, we can numerically fit for the first few $c_n$.\footnote{In principle we could also have explicitly numerically estimated the overlap integral $\int_{-1/2}^{1/2}dx \int_{\sqrt{1-x^2}}^{\infty} dy \,y^{-2}Z^{(c=2)}(\tau;\rho,\sigma) \nu_n(\tau)$ for fixed $\rho$ and $\sigma$, and found the $c_n$'s this way. In practice, however, we found that restricting to a specific spin, plugging in various values of $\tau$, $\rho$, and $\sigma$ and finding the best fit for the $c_n$'s (after cutting off the sum in $n$) led to far more accurate numerics. For simplicity, we restricted ourselves to parity-invariant theories, so we did not obtain numerical data on $c^-_n$.} We find the following values:
\begin{align} 
&c^+_1 \sim  7.99999999911,~~ c^+_2 \sim 8.0000000011, ~~ c^+_3 \sim 7.999999916,~~ c^+_4 \sim 8.00000043,\nn\\
&c^+_5 \sim 8.000013,~~ c^+_6 \sim 8.016, ~~ c^+_7\sim 8.00033, ~~ c^+_8 \sim 7.9953
\label{eq:8sgalore}
\end{align} 
with higher values of $n$ being less numerically stable. Our numerical results (\ref{eq:8sgalore}) are indeed consistent with the result
\be
c^+_n = 8.
\ee 

\subsubsection{Decompactification loci of \texorpdfstring{$c=3$}{c3} and \texorpdfstring{$c=4$}{c4}}

In \cite{Obers:1999um, Angelantonj:2011br}, an efficient method was described to perform spectral decomposition on the product locus of Narain moduli $T^d \times S^1$ where the radius of the $S^1$ is taken very large. Here we will focus on the case $d=2$. In Appendix \ref{sec:decomp}, we describe this method in detail; here we simply state the result. If we take $\rho, \sigma$ to be the moduli of the $T^2$ theory, and $r_3$ to be the radius of the $S^1$, then at large $r_3$ (with all other parameters fixed), we get
\begin{align}
Z^{(c=3)}(\tau; &\rho, \sigma, r_3) \simeq E_{3\o2}(\tau) + r_3 \(\frac 6\pi \log(r_3)+ \widehat E_1(\rho) + \widehat E_1(\sigma) + \beta \)
\nn\\& + \frac{1}{4\pi i} \int_{\re s = \half}ds\, E_s(\tau) \frac{ r_3 2 \Lambda^2(s) E_s(\sigma) E_s(\rho)    + 2 \Lambda(s)\Lambda(-s) r_3^{2s+1} + 2 \Lambda(1-s) \Lambda(s-1) r_3^{3-2s} }{ \Lambda(1-s)} \nn\\
&+ \sum_{\epsilon=\pm}\sum_{n=1}^{\infty} \nu^\epsilon_n(\tau) \frac{(Z^{(c=3)},\nu^\epsilon_n)}{(\nu^\epsilon_n, \nu^\epsilon_n)}
\label{eq:c3bla}
\end{align}
where $\simeq$ means equals to all orders in $1/r_3$ (i.e. the error is non-perturbative in $1/r_3$). In (\ref{eq:c3bla}), $E_{3\o2}(\tau)$ is the non-holomorphic Eisenstein series at $s=3/2$, and the constant term is inherited from the residue of the Eisenstein overlap at $s=1$
\begin{equation}
	\Res_{s=1}\left(R_s[Z^{(c=3)}]\right) \simeq r_3 \(\frac 6\pi \log(r_3)+ \widehat E_1(\rho) + \widehat E_1(\sigma) + \beta \)
\end{equation}
with
\begin{equation}
	\beta \coloneqq {6\over \pi}\left(\gamma_E+24\zeta'(-1)+\log (4\pi) -2\right) \approx -5.46576. 
\end{equation}

To fix the cusp form contributions, we again can numerically fit (\ref{eq:c3bla}) for various values of target space moduli. Remarkably they also seem to take a simple form at large $r_3$. We conjecture that at large $r_3$
\begin{align}
&Z^{(c=3)}(\tau; \rho, \sigma, r_3) \simeq E_{3\o2}(\tau) + r_3 \(\frac 6\pi \log(r_3)+ \widehat E_1(\rho) + \widehat E_1(\sigma) + \beta\)
\nn\\& + \frac{1}{4\pi i} \int_{\re s = \half}ds\,E_s(\tau) \frac{ r_3 2 \Lambda^2(s) E_s(\sigma) E_s(\rho)    + 2 \Lambda(s)\Lambda(-s) r_3^{2s+1} + 2 \Lambda(1-s) \Lambda(s-1) r_3^{3-2s} }{ \Lambda(1-s)} \nn\\
&+ 8r_3 \sum_{n=1}^{\infty} \frac{\nu^+_n(\tau) \nu^+_n(\rho) \nu^+_n(\sigma)}{(\nu^+_n,\nu^+_n)} - 8 i r_3 \sum_{n=1}^{\infty} \frac{\nu^-_n(\tau) \nu^-_n(\rho) \nu^-_n(\sigma)}{(\nu^-_n,\nu^-_n)}.
\label{eq:c3speciallocusts}
\end{align} 
Interestingly in (\ref{eq:c3speciallocusts}) we see that for $c=3$, unlike for $c=2$, there are special points in the moduli space (for instance, very large $r_3$) where the cusp forms contribute numerically a much larger contribution to the partition function than the average $E_{3\o2}(\tau)$. For the $c=2$ partition function (\ref{eq:c=2Spectral}) this is not the case due to the boundedness of the cusp forms. 

In Appendix \ref{sec:decomp} we also describe a generalization of the method introduced in \cite{Obers:1999um, Angelantonj:2011br} to consider the spectral decomposition of Narain partition functions on the product locus $T^d \times T^D$, where we take the volume of the $T^d$ to be parametrically large. Here we will focus on the case $d = D = 2$, which is a sublocus of the $c=4$ Narain moduli space. We will denote the moduli of the individual tori as $\rho^{(i)}$ and $\sigma^{(i)}$ for $i=1,2$, and will consider the limit in which $\rho_2^{(1)} = \im(\rho^{(1)})$ is taken very large. In this limit, up to non-perturbative corrections in $\rho_2^{(1)}$ we find
\begin{equation}
\begin{aligned}
	Z^{(c=4)}(\tau;m) &\simeq  \, E_2(\tau) +  \rho_2^{(1)}\left(\widehat E_1(\sigma^{(1)})+\widehat E_1(\rho^{(2)})+\widehat E_1(\sigma^{(2)})+{3\over \pi}\log\rho_2^{(1)} + \gamma\right)\\
	& \, + {1\over 4\pi i}\int_{\re s=\half}ds\, 2\rho_2^{(1)}\bigg[ {\Lambda(s)^2\over\Lambda(1-s)}E_s(\rho^{(2)})E_s(\sigma^{(2)}) + \left(\rho_2^{(1)}\right)^{1-s}\Lambda(s-1)E_{s-1}(\sigma^{(1)}) \\
	& \, \quad\quad\quad\quad\quad\quad\quad\quad\quad + \left(\rho_2^{(1)}\right)^s{\Lambda(s)\Lambda(s+1)\over\Lambda(1-s)}E_{s+1}(\sigma^{(1)}) \bigg]E_s(\tau)\\
	& \, + 8\rho_2^{(1)}\sum_{n=1}^\infty{\nu^+_n(\rho^{(2)})\nu^+_n(\sigma^{(2)})\over (\nu^+_n,\nu^+_n)}\nu^+_n(\tau) - 8i\rho_2^{(1)}\sum_{n=1}^\infty{\nu^-_n(\rho^{(2)})\nu^-_n(\sigma^{(2)})\over (\nu^-_n,\nu^-_n)}\nu^-_n(\tau) ,
	\label{eq:c4wooo}
\end{aligned}
\end{equation}
with
\begin{equation}
	\gamma \coloneqq
	{6\over \pi}(\gamma_E + 2\log(4\pi) + 36 \zeta'(-1)-3)\approx -6.3329.
	\label{eq:gammadeff}
\end{equation}

\subsubsection{\texorpdfstring{$c>2$}{cg2}}

We are now in a position to study the spectral decomposition of the partition functions of Narain lattice CFTs with $c>2$. As in previous subsections, we begin with the Eisenstein overlap, whose residue at $s=1$ gives the constant term, followed by the cusp form overlap. 

From (\ref{eq:DIFFEQ}), we see the RS transform of the Narain partition functions is an eigenfunction of the target space Laplacian $\Delta_{\mathcal{M}_c}$
\begin{equation}
	\Delta_{\mathcal{M}_c}R_s[Z^{(c)}] = \left({c\over 2}-s\right)\left({c\over 2}-1+s\right)R_s[Z^{(c)}].
\end{equation}
That the lattice theta function provides a pairing between eigenfunctions of the worldsheet and target space Laplacians is reminiscent of a concept in the math literature known as the ``theta correspondence" \cite{Deitmar_1991}. The RS transform is straightforwardly computed to be \cite{Angelantonj:2011br}
\begin{equation}
\begin{aligned}
	R_s[Z^{(c)}] 
	&= \int_0^\infty dy \, y^{s+{c\over 2}-2}\sideset{}{'}\sum_{n,w\in\ZZ^c}\delta_{n\cdot w,0}e^{-\pi y M_{n,w}(m)^2}\\
	&= \pi^{1-s-{c\over 2}}\Gamma\left(s+{c\over 2}-1\right)\sideset{}{'}\sum_{n,w\in\mathbb{Z}^c} {\delta_{n\cdot w,0}\over (M_{n,w}(m)^2)^{s+{c\over 2}-1}}\\
	&= \pi^{1-s-{c\over 2}}\Gamma\left(s+{c\over 2}-1\right) \mathcal{E}^c_{s+{c\over 2}-1}(m)
\end{aligned}
\end{equation}
where $m$ represents the moduli, and we have assumed $\re s > 1-{c\over 2}$ in order to exchange the sum and the integral. In \cite{Angelantonj:2011br}, $\mathcal{E}^c_s$ is referred to as a ``constrained Epstein zeta series''
\begin{equation}
	\mathcal{E}^c_{s}(m) \coloneqq \sideset{}{'}\sum_{n,w\in\mathbb{Z}^c}{\delta_{n\cdot w,0}\over M_{n,w}(m)^{2s}},
	\label{eq:constrainedepsteindef}
\end{equation}
which converges for $\re s > c-1$.\footnote{In e.g. \cite{Angelantonj:2011br}, it was written that convergence requires $\re s > c$, but it seems that the weaker condition $\re s > c - 1$ is enough for convergence. For $c > 2$, the density of states of scalar primary operators grows at large $\Delta$ as $\rho(\Delta) \sim \frac{(2\pi)^{c-1}\Lambda\(\frac{c-1}2\)}{\Gamma(c-1)\Lambda\(\frac c2\)} \Delta^{c-2}$, so for convergence of (\ref{eq:constrainedepsteindef}), we require that $c-2-\re s < -1$, or equivalently $\re s > c -1$. We are grateful to Cyuan-Han Chang for pointing this out to us.} It admits a meromorphic continuation to the entire $s$ plane and satisfies a functional equation, both inherited from the Eisenstein series: 
\begin{equation}
	\pi^{-s}\Gamma(s)\Lambda\left(s+1-{c\over 2}\right)\mathcal{E}^c_s(m) = \pi^{-(c-1-s)}\Gamma(c-1-s)\Lambda\left({c\over 2}-s\right)\mathcal{E}^c_{c-1-s}(m).
\end{equation}
The constant part of the spectral decomposition is extracted from the residue of $R_{1-s}[Z^{(c)}]$ at $s=1$, cf. \eqr{avgrs}. As far as the cusp forms, we do not have analytic control over their inner product with the Narain partition functions. However, we do know that they must be eigenfunctions of the target space Laplacian, in particular,
\begin{equation}
	\Delta_{\mathcal{M}_c}(Z^{(c)},\nu_n) = \left({(c-1)^2\over 4}+R_n^2\right)(Z^{(c)},\nu_n).
\end{equation}
Moreover from the argument in Appendix \ref{app:eisensteinCG}, we know that the inner products $(Z^{(c)},\nu_n)$ are generically non-zero. 
Altogether, therefore, the spectral decomposition of the partition function of a general Narain lattice CFT with $c>2$ takes the following form
\begin{empheq}[box=\fbox]{align}\label{eq:higherCSpectral}
	Z^{(c)}(\tau;m) &= E_{c\over 2}(\tau) + {1\over 4\pi i}\int_{\re s = \half} ds\, \pi^{s-{c\over 2}}\Gamma\left({c\over 2}-s\right)\mathcal{E}^c_{{c\over 2}-s}(m)E_s(\tau)\nonumber\\
	& + {3\over \pi}\pi^{1-{c\over 2}}\Gamma\left({c\over 2}-1\right)\mathcal{E}^c_{{c\over 2}-1}(m) + \sum_{n=1}^\infty \frac{(Z^{(c)},\nu_n)}{(\nu_n, \nu_n)}\nu_n(\tau).
\end{empheq}

Let us emphasize one important feature here. The Eisenstein series $E_{c\over 2}(\tau)$, present in (\ref{eq:higherCSpectral}) because the Narain partition functions are not square integrable (due to the identity operator), has recently been shown to have an elegant interpretation as the Narain partition function \emph{averaged} over the moduli space $\mathcal{M}_c$ with the measure inherited from the Zamolodchikov metric \cite{Afkhami-Jeddi:2020ezh,Maloney:2020nni}. So the other terms on the right hand side of (\ref{eq:higherCSpectral}) admit a natural interpretation as the deviation from the average. (We remind the reader that this average is only well-defined for $c>2$ \cite{Afkhami-Jeddi:2020ezh,Maloney:2020nni}.) We will say more about this interpretation in Section \ref{sec4}.

\subsection{On the optimal gap in the scalar sector}

Before moving on, we make a brief remark on how our methods may be useful in determining the largest possible gap to non-vacuum primary operators in Narain CFTs.

In the spectral representations of the Narain partition functions (\ref{eq:c=1Spectral}), (\ref{eq:c=2Spectral}), (\ref{eq:higherCSpectral}), modular invariance is completely manifest but unitarity (positivity of the spectrum), compactness (discreteness of the spectrum) and integrality (degeneracies are integers) are obscured. This is in contrast to standard approaches to the modular bootstrap, where positivity is manifest, but modular invariance, compactness\footnote{Although in numerical approaches to the modular bootstrap one typically demands positivity of the action of the linear functional on the contribution of the vacuum module to the modular crossing equation, in practice \cite{Keller:2012mr,Friedan:2013cba,Collier:2016cls} it is often quite difficult to rule out noncompact solutions to the modular crossing equations that are spurious for the problems of interest. A typical example is the question of the optimal upper bound on the gap in the spectrum of twists in a compact 2d CFT. There is a simple argument \cite{Collier:2016cls,Afkhami-Jeddi:2017idc} that shows the twist gap cannot exceed ${c-1\over 12}$, which is saturated by the noncompact Liouville CFT. Numerical approaches cannot improve on this bound without a novel mechanism demanding the presence of a normalizable ground state and/or a discrete spectrum of local operators (see \cite{Collier:2016cls, Benjamin:2019stq} for discussions on this).} and integrality are not. 

A simple example of a problem for which the technology developed in this paper may be suitable is the question of the optimal upper bound on the gap in the spectrum of scalar $U(1)^c\times U(1)^c$ primary operators, particularly in the large $c$ limit. The latter is a problem that evades even state-of-the-art numerical modular bootstrap analyses \cite{Afkhami-Jeddi:2019zci,Hartman:2019pcd,Afkhami-Jeddi:2020hde}. For the purposes of attacking this question, it is convenient to introduce a counting function $N_j(\Delta)$
\begin{equation}
	N_j(\Delta) \coloneqq \int_0^\Delta d\Delta'\, \left(\rho_j(\Delta')-\delta_{j,0}\delta(\Delta')\right)
\end{equation}
that enumerates the total number of spin-$j$ primaries with dimension less than or equal to $\Delta$. In particular, the scalar counting function $N_0(\Delta)$ is dictated entirely in terms of the overlap of the Narain partition function with the real analytic Eisenstein series, as the cusp forms have no scalar Fourier mode. From (\ref{eq:higherCSpectral}) we can read off the counting function by integrating the scalar part, which gives:
\begin{equation}
\begin{aligned}\label{eq:scalarCountingFunction}
	N_0(\Delta) =& \, {2\pi^c\zeta(c-1)\over(c-1)\Gamma\left({c\over 2}\right)^2\zeta(c)}\Delta^{c-1} + {3\over \pi}{(2\pi\Delta)^{c\over 2}\over \Gamma\left({c\over 2}+1\right)}f_c(1)\\
	& \, + {1\over 4\pi i}\int_{\re s=\half}\left[{(2\pi\Delta)^{{c\over 2}-s}\over\Gamma\left({c\over 2}-s+1\right)}+{\Lambda(1-s)\over\Lambda(s)}{(2\pi\Delta)^{{c\over 2}-1 + s}\over\Gamma\left({c\over 2}+s\right)}\right]f_c(s),
\end{aligned}
\end{equation}
where
\begin{equation}
	f_c(s) \coloneqq (Z^{(c)},E_s).
\end{equation}

The existence of a gap in the spectrum implies that $f_c(s)$ has certain analytic properties in the complex $s$-plane. For sufficiently small $\Delta$, we should be able to evaluate the counting function (\ref{eq:scalarCountingFunction}) by deforming the $s$ contour away from the critical strip to infinity to get zero. For example, in order to cancel the contribution of the first term in (\ref{eq:scalarCountingFunction}), representing the continuous density of states of the average, we know that $f_c(s)$ must have a simple pole at $s={c\over 2}$ with unit residue
\begin{equation}
	\Res_{s={c\over 2}}f_c(s) = 1.
\end{equation} 
Discreteness of the spectrum implies that $\frac{\Lambda(1-s)}{\Lambda(s)} f_c(s)$ must not have any other poles in the half-plane to the right of the critical line, as the existence of such a pole would lead to a continuous contribution to the counting function $N_0$. The existence of a scalar gap also implies that $f_c(s)$ has certain asymptotic properties as $s$ is taken to infinity, for example in the right half-plane. If we define $g_c(s)$ by rescaling
\begin{equation}
	g_c(s) \coloneqq {\Lambda(1-s)\over \Lambda(s)}{1\over \pi^{-({c\over 2}-1+s)}\Gamma\left({c\over 2}-1+s\right)}f_c(s)
\end{equation}
then if the scalar gap is $\Delta_*$, we learn that $g_c(s)$ must fall off as $s \to\infty$ in the right half $s$-plane as follows
\begin{equation}
	g_c(s) \stackrel{s\to\infty}{\sim} (2\Delta_*)^{-s}.
	\label{eq:gcsgap}
\end{equation}
If this is the case, then for any $\Delta<\Delta_*$ we can deform the $s$-contour and get $N_0(\Delta<\Delta_*)=0$. Then the question is: for Narain CFTs at fixed central charge, how large can $\Delta_*$ be?

To get a feeling for how this works, it is instructive to consider some cases for which we already know the optimal bound on the gap. For example, in the $c=1$ case, we know from (\ref{eq:c=1EisensteinOverlap}) that
\begin{equation}
	g_1(s) = 2\zeta(2s-1)\left(r^{2s-1}+r^{1-2s}\right) \stackrel{s\to\infty}{\approx} 2\left(r^{2s-1}+r^{1-2s}\right).
\label{eq:gcc1}
\end{equation}
The bound on the gap is optimized when this quantity is minimized, corresponding to $r=1$. For this choice of $r$, $g_c$ is asymptotically constant as $s$ is taken to infinity, corresponding to an optimal gap of $\Delta_* = \half$, which is indeed saturated by the $c=1$ free boson at the self-dual radius.\foot{More generally, let us consider a $T$-duality fundamental domain that maximizes $r$, namely $r\in [1, \infty)$. Then (\ref{eq:gcc1}) at large $s$ is $g_1(s) \approx r^{2s}$ which when compared with (\ref{eq:gcsgap}) gives $\Delta_*^{(c=1)} = 1/2r^2$.} Similarly, in the $c=2$ case, we know from (\ref{eq:c=2InnerProduct}) that
\begin{equation}
	g_2(s) = 2\zeta(2s)E_s(\rho)E_s(\sigma).
\end{equation}
Let us assume $\rho, \sigma$ are in the usual fundamental domain of $\mathbb{H}/SL(2,\mathbb Z)$ (which maximizes the imaginary part over the $SL(2,\mathbb Z)$ orbits). Then as $\re s$ is taken to infinity, $g_2$ is approximated by
\begin{equation}
	g_2(s) \approx 2(\rho_2\sigma_2)^s.
\end{equation}
Comparing to (\ref{eq:gcsgap}), this tells us that in a given Narain lattice CFT with $T^2$ target space,
\begin{equation}
	\Delta_*^{(c=2)} = {1\over 2\rho_2\sigma_2},
\end{equation} 
provided $-\frac12 < \rho_1, \sigma_1 \leq \frac12$ and $|\rho|, |\sigma| \geq 1$. Since $\rho$ and $\sigma$ are valued in the fundamental domain, we know $\rho_2,\sigma_2 \geq {\sqrt{3}\over 2}$. Therefore the optimal gap is $\Delta_* = {2\over 3}$. This is saturated by the $SU(3)_1$ WZW model \cite{Collier:2016cls,Afkhami-Jeddi:2020ezh}, corresponding to the complexified K\"ahler and complex structure moduli at the $\mathbb{Z}_3$ point, $\rho = \sigma = e^{2\pi i\over 3}$.

More generally for $c>2$, we can make statements about the scalar gap as a function of the $O(c,c;\mathbb{Z}) \backslash O(c,c; \mathbb R)/O(c)\times O(c)$ moduli. We conjecture that $g_c(s)$ at large $\re s$ will take the simple analytic form of $\frac1{2(G^{-1})_{cc}}$ when written in variables that minimize $(G^{-1})_{cc}$ (i.e. the $T$-duality fundamental domain is chosen to minimize $(G^{-1})_{cc}$). The scalar gap would then be optimized at a cusp point in the target space moduli, corresponding to a specific rational CFT. It would be very interesting to have more explicit expressions at higher $c$, and to compare to candidate optimal Narain theories found by scanning moduli space in Table 3.1 of \cite{Afkhami-Jeddi:2020ezh}.

\section{Application to general CFTs}\label{sec4}

We now explore the role of spectral decomposition in general CFTs. 

\subsection{Rendering the partition function square-integrable}
 
Unlike the Narain case, there is an immediate obstruction to applying the spectral decomposition to generic, i.e. irrational, CFT partition functions. For a CFT with extended chiral algebra $\mathcal{A}$, we may define the quantity $c_{\rm currents}\geq 1$ as the effective central charge in the high-temperature asymptotics of the vacuum character of $\mathcal{A}$:
\begin{equation}\label{ccurr}
\log\chi_{\rm vac}^{\mathcal{A}}(y \rar 0) \sim {i\pi\over 12\tau} c_{\rm currents}.
\end{equation}
This quantity counts the number of currents modulo non-trivial null states. Due to the identity operator, the primary partition function of the CFT near the cusp behaves as (suppressing a power-law prefactor) 
\e{zpcusp}{Z_p(y\rar\infty) \sim q^{-{c-c_{\rm currents}\o 12}}.}
Whereas Narain CFTs obey $c=c_{\rm currents}$, a generic CFT obeys $c>c_{\rm currents}$, so \eqr{zpcusp} diverges exponentially. Indeed, any {\it light} primary operator, defined by the condition
\be
h + \hb \leq \frac{c-c_{\rm currents}}{12},
\label{eq:inequalityasdf}
\ee
renders $Z_p(\t)\notin L^2(\mathcal{F})$; moreover, those contributions strictly below the threshold are not even of ``slow growth'' near the cusp $y\rightarrow \infty$, in the sense of Zagier \cite{zbMATH03796039}. 

We henceforth take the CFTs in question to be compact, with Virasoro symmetry alone and $c>c_{\rm currents}=1$, leaving implicit the straightforward extension to larger chiral algebras. The primary partition function is
\be
Z_p(\t) = y^{1/2} |\eta(\tau)|^2 Z(\tau).
\label{eq:reducedpf}
\ee
It will be useful to introduce a primary partition function over light operators only:
\e{Zlight}{Z_L(\t) \coloneqq y^{1/2}  \sum_{\substack{h,\hb \\ h + \bar h\, \leq\, {c-1\over 12}}} q^{h-{c-1\o24}}\bar q^{\bar h-{c-1\o24}}}
where each term is the contribution of a Virasoro primary operator of weight $(h,\hb)$. At finite $c$, there are a finite number of such contributions. As it stands, (\ref{eq:reducedpf}) does not have a Roelcke-Selberg spectral decomposition because  $Z_L(\t) \neq 0$, due to the vacuum and any other light operators.

However, let us view this problem slightly differently. We can subtract all light primary operators in a modular-invariant way, such that the resulting function {\it will} be in $L^2(\mathcal{F})$. In particular, let us define 
\e{Zspec}{Z_{\rm spec}(\t) \coloneqq Z_p(\t) - \widehat Z_{L}(\t).}
The quantity $\widehat Z_L(\t)$ is defined to be a {\it modular completion} of $Z_L(\t)$. That is, given $Z_L(\t)$, one adds states to make it modular-invariant, 
\begin{equation}
\widehat Z_L(\g\t) = \widehat Z_L(\t)\,, \quad \g\in SL(2,\ZZ),
\end{equation}
but without adding any new light states, 
\e{}{\widehat Z_L(\t) - Z_L(\t) \rightarrow 0 ~~\text{as}~~ y\rightarrow \i.}
Subtracting this quantity from $Z_p(\t)$ as in \eqr{Zspec}, the resulting $Z_{\rm spec}(\t)$ {\it is} square-integrable, and therefore admits a spectral decomposition of Roelcke-Selberg type, cf. \eqr{eq:RoelckeSelberg}.

The modular completion $\widehat Z_L(\t)$ of a given $Z_L(\t)$ is not unique. Indeed, mathematically, there are an infinite number of ways to modular complete which differ by ``cuspidal functions,'' i.e. modular-invariant functions which vanish at $y \rightarrow 0$. One particular modular completion that seems privileged, and provides insight into the meaning of $Z_{\rm spec}(\t)$, is to replace each light state by its $PSL(2,\ZZ)$ Poincar\'e sum:
\e{ZLpoinc}{\widehat Z_L(\t) = \sum_{\substack{h,\hb \\ h + \bar h \, \leq\, {c-1\over 12}}} \sum_{\gamma\in\Gamma_\infty\backslash PSL(2,\ZZ)} \text{Im}(\g\t)^{1/2} q_\g^{h-{c-1\o24}}\overline q_\g^{\hb-{c-1\o24}}} 
where $q_\g \coloneqq \exp(2\pi i \g \t)$ and $\overline q_\g \coloneqq \exp(-2\pi i \g \bar \t)$. The Poincar\'e sum of each term adds, upon regularization \cite{Keller:2014xba, Benjamin:2020mfz}, states with $h \geq {c-1\o 24}$ and $\hb \geq {c-1\o 24}$ to the original state. See appendix \ref{app:generalizedEisenstein} for more details on the generalized Eisenstein series that appear when considering the modular completion via Poincar\'e series of light characters. Part of the appeal of \eqr{ZLpoinc} is that at large central charge, the Poincar\'e sum is quite natural from the point of view of quantum gravity in AdS$_3$. In particular, given a light state, a Poincar\'e sum generates contributions to the heavy spectral density that qualitatively reproduce the behavior of AdS$_3$ gravity in the small-$G_N$ expansion. That Poincar\'e sums ``look'' gravitational has long been observed \cite{Dijkgraaf:2000fq, Maloney:2007ud, Castro:2011zq, Keller:2014xba}; recent work has augmented this point of view \cite{Alday:2019vdr, Benjamin:2020mfz,Afkhami-Jeddi:2020ezh,Maloney:2020nni,Cotler:2020ugk,Perez:2020klz, Maxfield:2020ale, Alday:2020qkm,Dymarsky:2020pzc,Meruliya:2021utr,Benjamin:2021wzr,Datta:2021ftn, Meruliya:2021lul,Ashwinkumar:2021kav,Dong:2021wot} in part due to the possibility that AdS$_3$/CFT$_2$ may be understood as involving ensemble averages over CFTs, which assigns a physical interpretation to a continuous density of states. 

With that said, there is no clear-cut candidate for a canonical modular completion. For example, a very interesting method of modular completion that leads to a slightly different heavy spectrum than that of the Poincar\'e sum, based on ideas used in Rademacher expansions, was recently put forth in \cite{Alday:2019vdr}. It is not obvious to us if the modular completion using the Poincar\'e or Rademacher formalism is a more ``natural" candidate. One may instead wish for a modular completion of the contribution of a given light state to the partition function that preserves discreteness of the accompanying heavy spectrum. However, we are not aware of such a modular completion that also retains square-integrability.\foot{For example, iteratively subtracting powers of the modular $j$-function from $Z(\t)$ to eliminate the singularity at $y\rar\infty$ necessarily involves negative powers, which in turn introduces new singularities elsewhere due to the zero of the $j$-function on the unit $\t$-circle.}

\subsection{Interpretation}

We are led to suggest the following perspective on the role of spectral decomposition in general 2d CFTs: \textit{given a light spectrum, $Z_{\rm spec}(\t)$ computes a deviation from average of the primary partition function $Z_p(\t)$, where the role of the ``average'' is played by $\widehat Z_L(\t)$.}

This dovetails with our previous treatment of Narain CFTs, where ``average'' is taken to mean the literal ensemble average over Narain moduli space. Recognizing $E_{c\over 2}(\t)$ as the average primary partition function \eqr{eq:stripoffcrap} of the $U(1)^c \times U(1)^c$ Narain CFTs, we may rewrite \eqr{eq:higherCSpectral} as
\es{eq:individualNarain}{Z^{(c)}(\tau;m) - \langle Z^{(c)}(\tau;m) \rangle =& \, {1\over 4\pi i}\int_{\re s = \half} ds\, \pi^{s-{c\over 2}}\Gamma\left({c\over 2}-s\right)\mathcal{E}^c_{{c\over 2}-s}(m)E_s(\tau)\\
	& \, + {3\over \pi}\pi^{1-{c\over 2}}\Gamma\left({c\over 2}-1\right)\mathcal{E}^c_{{c\over 2}-1}(m) + \sum_{n=1}^\infty (Z^{(c)},\nu_n)\nu_n(\tau).}
The right-hand side is $Z_{\rm spec}^{(c)}(\t)$, obeying $\langle Z_{\rm spec}^{(c)}(\t)\rangle =0$. So this is precisely of the form \eqr{Zspec} with 
\e{}{\widehat Z_L(\t)=\langle Z^{(c)}(\tau;m) \rangle=E_{c\over 2}(\t)}
where $E_{c\over 2}(\t)$ is, we emphasize, the Poincar\'e sum of the identity operator. 

From this point of view, it is rather interesting that $E_{c\over 2}(\t)$ is the average Narain partition function. Suppose one were to average over Narain moduli space with a measure {\it different} from the Zamolodchikov metric. Then $\widehat Z_L(\t)$ and $\langle Z^{(c)}(\t) \rangle$ would not necessarily be equal. This gives a novel perspective on why the Zamolodchikov metric is privileged, and motivates the Poincar\'e sum as a natural modular completion: the average over moduli space, the Rankin-Selberg method, and the Poincar\'e sum of the identity operator all coincide.

In the general holographic context with Virasoro symmetry, we view $\widehat Z_L(\t)$ as an ``average'' in the sense of {\it universality}. For the modular completion via Poincar\'e sum, for example, $\widehat Z_L(\t)$ captures universal contributions of light matter to the black hole spectrum. This takes the form of the Cardy entropy plus an infinite series of corrections from the full family of $SL(2,\ZZ)$ black holes \cite{Maldacena:1998bw}. It encodes the semiclassical gravity approximation of the microstate counts of those black holes that form via collapse of light matter. The quantity $Z_{\rm spec}(\t)$ therefore accesses the microstructure that lies beneath the coarse-grained approximation, by subtracting these contributions from the full partition function $Z_p(\t)$. Note that, unlike the Narain case reviewed above, we do not know how to ensemble average general Virasoro primary partition functions in the absence of moduli, and different modular completions $\widehat Z_L(\t)$ give rise to different $Z_{\rm spec}(\t)$ for a given theory. The suggestion of the Narain case is this: if there exists a formal notion of ``ensemble averaging'' over Virasoro CFTs without moduli, a natural definition would be one for which $\langle Z_{\rm spec}(\t) \rangle =0$.

If $\widehat Z_L(\t)$ is to be interpreted in the above way, it is natural to expect that it should satisfy some physical constraints. For instance, in order to interpret $\widehat Z_L(\t)$ as capturing universality of a (putative) consistent gravitational theory in AdS$_3$, one would like its inverse Laplace transform to be positive-definite.\foot{This need only hold for those light spectra consistent with all CFT axioms, in the spirit of the modular bootstrap; of course, it remains an open problem to determine the constraints on such spectra.} As currently formulated, this may not hold in all regions of $(h,\hb)$ because the present application of harmonic analysis is firmly {\it Euclidean}: CFT partition functions are finite everywhere in $\mathcal{F}$ away from the cusp at $y\rar\infty$, so square-integrability is sensitive to the spectrum of dimensions, $\Delta$, but not to the spectrum of twists, $t \coloneqq 2\,\text{min}(h,\hb)$. As such, the modular completion $\widehat Z_L(\t)$ may not capture certain universal properties of twist spectra. For example, if one chooses to compute $\widehat Z_L(\t)$ via Poincar\'e sum, the result will be positive-definite only if $Z_L(\t)$ includes an operator with $t \leq \frac{c-1}{16}$ \cite{Benjamin:2019stq}. Also, independently, if a CFT includes an operator lying strictly below this bound, it also includes infinitely many other operators, organized into Regge trajectories of asymptotically large spin, with asymptotic twist less than $\frac{c-1}{12}$ \cite{Kusuki:2018wpa, Collier:2018exn}. Neither of these phenomena would be fully captured by the Poincar\'e sum $\widehat Z_L(\t)$, though other modular completions might plausibly do better. This is not a contradiction, but it does motivate the extension of $SL(2,\ZZ)$ harmonic analysis to {\it Lorentzian} regimes.

\subsection{On half-wormholes in 2d CFT}

The above discussion suggests an analogy between structures in 2d CFT partition functions and recent observations about saddle points in holography. It has been shown that in a toy version of the SYK model, the partition function contains novel saddle points, dubbed ``half-wormholes'' \cite{Saad:2021rcu}. The general picture of \cite{Saad:2021rcu, shenker} is that the partition function of the large $N$ SYK model \emph{without} disorder average is a sum of two saddle points: a disk-type saddle point, dual to a smooth black hole geometry in the bulk, and a half-wormhole, a ``noisy'' saddle point dual to some bulk geometry which gives small corrections to the black hole.

We can identify parallel structures in 2d CFT partition functions. In particular, at large central charge, $\widehat Z_L(\t)$ captures the configurations continuously connected to saddle points, while $Z_{\rm spec}(\t)$ is a certain 2d analog of the half-wormhole. This follows rather naturally from our earlier discussion, since $\widehat Z_L(\t)$ captures black hole universality of semiclassical gravity, while $Z_{\rm spec}(\t)$ encodes small corrections, i.e. the ``noise.'' The total partition function is the sum of the two. 

This picture becomes especially sharp for the Narain theories whose partition functions are written, for example, in (\ref{eq:higherCSpectral}).  In this case $\widehat Z_L(\t)$ is precisely the averaged partition function $E_{c\over 2}(\tau)$, while the remaining terms describe  $Z_{\rm spec}(\t)$.  As is apparent from (\ref{eq:higherCSpectral}), many of the terms in $Z_{\rm spec}(\t)$ can be written in a way which suggests their origin as a sum over geometries.  For example, the Eisenstein series $E_s(\tau)$ is a Poincare series (see equation (\ref{eq:eisensteinDefinition}))  which can be interpreted as coming from a sum over geometries (handlebodies) labelled by $\Gamma_\infty \backslash SL(2,\ZZ)$ in the usual way.  This suggests, therefore, that we interpret these terms as the analog of half-wormhole contributions in the theory of gravity dual to the Narain ensemble.  It would be interesting, of course, to give a more explicit bulk interpretation of these contributions.

One crucial feature of the half-wormhole solutions is that they restore factorization.  
In particular, in the computation of the two-boundary path integral of individual instantiations in the model of  \cite{Saad:2021rcu} (see also \cite{Mukhametzhanov:2021nea}), i.e. before disorder average, the half-wormhole terms combine with the wormhole (present in the disorder-averaged SYK model) in such a way as to restore factorization of the square of the partition function. We can, at least in principle, explore this mechanism explicitly in Narain CFTs. We know from \cite{Maloney:2020nni} that the two-point function of the torus partition function averaged over Narain moduli space can be written in terms of a degree-two Eisenstein series (see \cite{Collier:2021rsn} for more details on the averaged two-point function)
\begin{equation}
	\langle Z^{(c)}(\tau_1)Z^{(c)}(\tau_2)\rangle = E_{c\over 2}^{(2)}(\Omega) = \sum_{\gamma\in P \backslash Sp(4,\mathbb{Z})}\left(\det\im\gamma\Omega\right)^{c\over 2} = \sum_{``(C,D) = 1"}{(y_1y_2)^{c\over 2}\over |\det\left(C\Omega+D\right)|^{c}},
\end{equation}
where $\Omega = \diag(\tau_1,\tau_2)$ is a diagonal element of the degree-two Siegel upper half-space, and the rest of the technical details of the above formula are unimportant for present purposes. A natural generalization of (\ref{eq:individualNarain}) would be something like 
\begin{equation}\label{eq:narainProduct}
	Z^{(c)}(\tau_1;m)Z^{(c)}(\tau_2;m) = \langle Z^{(c)}(\tau_1)Z^{(c)}(\tau_2)\rangle + Z_{\rm spec}^{(g=2)}(\t_1,\t_2;m).
\end{equation}
where $Z_{\rm spec}^{(g=2)}$ admits an $Sp(4,\ZZ)$ spectral decomposition. The left-hand side is computed by squaring (\ref{eq:higherCSpectral}), while the right-hand side is what is suggested by \cite{Saad:2021rcu} and our perspective on spectral decomposition, now at genus-two. It is a concrete problem to understand whether such an equality makes sense, and how to perform $Sp(4,\ZZ)$ harmonic analysis, given the expressions in Section \ref{sec:narain} of this paper (see \cite{Pioline:2014bra,Florakis:2016boz}, where the Rankin-Selberg method has been generalized to higher-genus modular functions).

\subsection{Spectral determinacy}\label{secspecdet}

The spectral methods herein shed light on the question of {\it spectral determinacy} in compact 2d CFT; that is, whether the complete spectrum is fully determined once part of it is fixed. Indeed, applying these methods to primary partition functions readily gives the following result:

\begin{quotation}
\noindent {\it The entire spectrum of a 2d CFT is uniquely fixed by the light spectrum, the scalar spectrum, and the spin $j=1$ spectrum.
}
\end{quotation}

\noindent This is depicted in a Chew-Frautschi plot in Figure \ref{specfig} for Virasoro CFTs, but the result applies for any extended chiral algebra, with ``light'' primaries defined by \eqr{eq:inequalityasdf}. The proof -- indeed, the algorithm for constructing the spectrum -- may be simply stated as follows:

{\bf i)} Given the light spectrum, form $Z_{\rm spec}(\t)$ defined in \eqr{Zspec}. Being square-integrable, $Z_{\rm spec}(\t)$ admits a spectral decomposition \eqr{eq:RoelckeSelberg}. 

{\bf ii)} The $j=0$ spectrum of $Z_{\rm spec}(\t)$ fixes the overlap $(Z_{\rm spec}, E_s)$, via the RS transform \eqr{eq:RankinSelberg}. This in turn fixes all higher-spin data coming from the Eisenstein part of $Z_{\rm spec}(\t)$. 
This step hinges on the fact that cusp forms have no scalar support, cf. (\ref{eq:nuexpansionasdf}). 

{\bf iii)} The $j=1$ spectrum of $Z_{\rm spec}(\t)$,  after subtracting the contribution from the Eisenstein part, fixes the overlap $(Z_{\rm spec},\nu_n)$, and hence the remaining cusp form contribution.\footnote{In order to read off the cusp form contribution, we must invert the modified Bessel function of the second kind. This is done using orthogonality relations given in e.g. equation (3.22) of \cite{Whittaker}. In general, if we define the subtracted spin-$j$ spectrum,
\be\label{reducedspinj}
Z_{\rm spec}^{\text{spin-}j}(y) \coloneqq \int_{-1/2}^{1/2} dx \, e^{-2\pi i j x} \(Z_{\rm{spec}}(\tau) - \frac{1}{4\pi i}\int_{\re s=\frac12} ds (Z_{\text{spec}}, E_s) E_s(\tau)\),
\ee
then we can read off the cusp form support by (assuming $a_j^{(n)} \neq 0$)
\be
\frac{(Z_{\text{spec}}, \nu_n)}{(\nu_n, \nu_n)} = - \frac{4 R_n \sinh(\pi R_n)}{\pi a_j^{(n)}} \lim_{\epsilon\rightarrow 0} \frac{1}{\log \epsilon} \int_{\epsilon}^{\infty} \frac{dy}{y^{3/2}} K_{i R_n}(2\pi jy) Z_{\rm spec}^{\text{spin-}j}(y).
\ee} 

This proof relies on an important assumption, which is that the cuspidal eigenspectrum is ``simple'': that is, the multiplicity of cusp forms with spectral parameter $R_n$ is no greater than one. This is unproven, but widely held to be true (e.g. \cite{cusp82,hejh,sque}). A direct implication of such non-degeneracy is that all cusp forms are not just eigenfunctions of the Laplacian, but also eigenfunctions of all Hecke operators $T_j$.\foot{The assumption of a simple eigenspectrum, or a restriction to Maass cusp forms that are eigenvalues of all Hecke operators (so-called ``Hecke-Maass forms''), is often an input in derivations of various theorems, see e.g. \cite{quesound,Ghosh_2013}. The best known bound on the multiplicity $d(R_n)$ is, up to an overall constant, $d(R_n) \lesssim R_n/\log R_n$ as $R_n\rar\i$, which is rather far from one \cite{ SLetter}.} Consequently, the spin $j=1$ coefficients $a_1^{(n)}$ of the cusp forms are all non-vanishing, the proof of which we review in Appendix \ref{subApp:cusp}. It is suggestive that our spectral determinacy result relies not just on the harmonic analysis of modular functions, but on the number-theoretic structure of these eigenfunctions related to Hecke operators.  We suspect that Hecke operators may have an important role to play in future studies of 2d CFT.

In fact, we can derive a somewhat more general result, subject to a further minor assumption: that the entire spectrum of a 2d CFT is uniquely fixed by the light spectrum, the scalar spectrum, and the spectrum of any fixed integer spin $j>0$.  The argument is exactly as given above, except that one now requires that the coefficients $a_j^{(n)}$ are all non-vanishing. When $j>1$ this cannot be easily proven, but it is certainly very plausible.  Indeed, as reviewed in the next section, the Fourier coefficients $a_j^{(n)}$ (for prime spin $j$) are essentially random real numbers distributed according to the semi-circle law.  Thus, although it is not rigorously proven that they are all non-zero, it is almost certainly true. 

It is interesting to contrast these results on spectral determinacy with those in narrower classes of CFTs. In holomorphic CFTs, the heavy spectrum $h>{c\o 24}$ is fully determined by its complement because the partition function is a weakly holomorphic modular form of $SL(2,\ZZ)$ \cite{rademacher1943expansion}. In rational CFTs, the twist spectrum $t> {c\o 12}$ is fully determined by its complement, this time due to properties of vector-valued modular forms \cite{Kaidi:2020ecu}. Note that spectral determinacy in $\Delta$ also applies to rational CFTs: performing the algorithm with respect to the full partition function $Z(\t)$ rather than to $Z_p(\t)$, whose form is obfuscated by non-trivial null states, yields essentially the same result, marginally weaker due to the lack of distinction between primaries and descendants. 

Placed in a wider CFT context, it is an interesting and curious feature of the above result that all spins $j>0$ are on equal footing (at least if the conjecture about the $a_j^{(n)}$ non-vanishing holds). The Lorentzian inversion formula and non-perturbative Regge bound of \cite{Caron-Huot:2017vep} imply that in $d$-dimensional CFTs, all primary operators of integer spin $j>0$ lie on analytic Regge trajectories. (For CFTs that are opaque in the sense of \cite{Caron-Huot:2020ouj}, the bound is instead $j>1$.) Their OPE data are subtly intertwined: one may not dial dimensions or structure constants of any finite number of operators at will. We are finding a more specific result for Virasoro primary degeneracies in 2d CFTs, at least once all light primaries are accounted for: not only are the spin $j>0$ degeneracies {\it intertwined}, but the degeneracies of all spins $j>0$ are entirely {\it determined} by those of a single spin (in addition to the scalar data used to fix the Eisenstein contribution, cf. \eqr{reducedspinj}). As for $d$-dimensional Regge trajectories, the scalar data is treated separately. It would be of great interest to understand 2d CFT spectral determinacy from a conformal Regge theory point of view. 

In fact, it may be the case that the nonzero spin data need not be specified as extra input. If two compact CFTs have identical light primaries and scalar primaries, then their difference of spectra is a discrete sum of delta functions which can be written entirely as a linear combination of Maass cusp forms. If one can show that this is impossible, i.e. that no linear combination of Maass cusp forms can give a function whose inverse Laplace transform is a sum of delta functions, this would complete the argument. Similar determinacy arguments emphasizing the strong constraints implied by compactness were given in \cite{Kaidi:2020ecu}. As we discuss in the next subsection, this impossibility is a reasonable expectation because cusp forms are, in a precise sense we will articulate in the next subsection, chaotic. We do not know how to prove the above statement and leave it for future study.\foot{An observation that might suggest an alternate version of spectral determinacy is that the logarithm of the partition function can be made square-integrable without subtracting a modular completion of light states, assuming the partition function is positive everywhere in the fundamental domain. In particular, $\log Z(\tau)$ has a linear divergence as $y \rightarrow \infty$, which can be subtracted off in a modular-invariant way by an appropriate factor multiplying $\widehat E_1(\tau)$:
\be
\log Z(\tau) - \frac{2\pi c}{12} \widehat E_1(\tau) \in L^2(\mathcal{F}).
\label{eq:logggg}
\ee
The meaning and utility of this are unclear. It may nevertheless be interesting to consider its spectral decomposition. Similarly, although the average of the $c=2$ Narain primary counting partition function diverges, the average of its logarithm converges. It is not clear to us how to compute this other than numerically. }

\subsection{On chaos and cusp forms}\label{subsec:chaosAndCusps}

Our discussion of $Z_{\rm spec}(\t)$ suggests that it is probing a certain fine structure of the spectrum of CFTs, the noise above the background of universal features. Here we will explore the idea that $Z_{\rm spec}(\t)$ --- more specifically, the presence or absence of Maass cusp forms --- is a probe of chaos in the spectrum of the CFT. Of course, that cusp forms participate in the spectral decomposition of the Narain partition functions (including at rational points in moduli space) prevents us from taking this idea literally as stated. Nevertheless, both the distribution of cusp form eigenvalues $R_n$, and the Fourier coefficients $a_j^{(n)}$ of a fixed cusp form $\nu_n$, are known to exhibit certain forms of chaos. This has been explored in the mathematics literature, see e.g. \cite{sarnak, PhysRevLett.69.2188,1993MaCom..61..245H,Steil:1994ue, Sarnak_1987}.

Let us first address the eigenvalues $R_n$. Unlike random matrix ensembles (GUE, GOE, etc.), the eigenvalues at asymptotically large $n$ obey a Poisson distribution \cite{sarnak, PhysRevLett.69.2188}. The eigenvalues are the (discrete part of the) quantum mechanical energy spectrum of a particle propagating on the fundamental domain $\mathcal{F}$ of $PSL(2,\mathbb{Z})$ with Hamiltonian given by the Laplacian $-\Delta_\tau$, a system that is known to be strongly chaotic. This is known as ``arithmetic chaos'' \cite{PhysRevLett.69.2188,sarnak,Bolte:1993ur}, terminology that is due to the existence of an infinite number of operators that commute with the Hamiltonian (the Hecke operators (\ref{eq:heckeDefinition})) and establish a deep connection with number theory.

The Fourier coefficients $a_j^{(n)}$ of the cusp forms themselves also exhibit properties related to chaos. As discussed in Appendix \ref{subApp:cusp}, the Fourier coefficients are the eigenvalues of the cusp forms under the action of the Hecke operators $T_j$, and the cusp forms are conventionally normalized so that the spin-one coefficient is unity, $a_1^{(n)} = 1$. Although this normalization of the cusp forms may seem contrived from a physical point of view, it turns out to be quite natural for a number of purposes. For example, the \textbf{Ramanujan-Petersson conjecture} (as stated e.g. in \cite{Terras_2013,Sarnak_1987}) states that the absolute value of the Fourier coefficients with prime $j$ are bounded
\begin{equation}
    |a^{(n)}_j| \le 2.
\end{equation}
It turns out that statistical properties of the Fourier coefficients of the cusp forms can also be naturally stated with this normalization. For example, if one considers the sequence of Fourier coefficients $\{a^{(n)}_j\}$ for \emph{fixed} prime $j$ ordered by increasing eigenvalue $R_n$, then Sarnak has shown that they are equidistributed with respect to the following measure \cite{Sarnak_1987}
\begin{equation}
    d\mu_j(t) = {(j+1)\sqrt{4-t^2}\over 2\pi \left[\left(j^{\frac12}+j^{-\frac12}\right)^2-t^2\right]}dt,~ |t|<2.
\end{equation}
Note that as $j\to \infty$, this measure is asymptotically Wigner's semicircle. Moreover, the \textbf{Sato-Tate conjecture} states that one can reverse the limits $j\to\infty$ and $n\to\infty$ and arrive at the same conclusion: for a \emph{fixed} cusp form, its prime Fourier coefficients are equidistributed with respect to a measure that is asymptotically Wigner's semicircle. This has been extensively checked numerically for the cusp forms with lowest-lying eigenvalues \cite{1993MaCom..61..245H,Steil:1994ue}.

It is tempting to speculate that this notion of chaos is related to spectral chaos in CFTs. It is however not clear how to make this relation precise. As shown previously, the Narain CFTs for $c>1$ all have cusp form support, though with a miraculously simple structure at e.g. $c=2$ (see (\ref{eq:eight})). We can also look at some examples of rational CFTs and see if there is any pattern in the behavior of their cusp form support. Let us compute the following quantity for the $(p,q)$ Virasoro minimal models:
\be
y^{1/2}|\eta(\t)|^2 Z_{(p,q)}(\t)
\ee
This is somewhat unnatural -- $|\eta(\t)|^2$ does not count Verma module descendants, due to the non-trivial null states -- but we proceed nevertheless. We now recall that partition functions for all Virasoro minimal models, diagonal or not, may be written as linear combinations of $Z^{(c=1)}(\t;r)$ of different radii \cite{DiFrancesco:1987gwq}. For example, diagonal invariants obey
\be
Z^{\rm diag}_{(p,q)}(\t) = {\frac12}\(Z^{(c=1)}\(\t;\sqrt{pq}\) - Z^{(c=1)}\(\t;\sqrt{p^{-1}q}\)\)
\ee
It follows from (\ref{eq:c=1Spectral}) that the above quantity does not have cusp forms in its spectral decomposition. The fact that the cusp forms do not show up for Virasoro minimal models seems to be an accident of these particular CFTs, however. For instance, we have verified numerically that $y|\eta(\tau)|^4$ times the partition function of the lowest unitary $W_3$ minimal model does have cusp forms in its decomposition. It remains unclear to us what the precise connection is between the Maass cusp form overlap of the partition function and quantum chaos.

\section*{Acknowledgments}
We are grateful to Fernando Alday, Cyuan-Han Chang, Tolya Dymarsky, Michael Green, Tom Hartman, Jeff Harvey, Peter Humphries, Yuya Kusuki, Ying-Hsuan Lin, Hirosi Ooguri, Boris Pioline, Sylvain Ribault, Steve Shenker, Andrew Sutherland, Yifan Wang, Xinan Zhou, and especially Herman Verlinde for very helpful discussions. NB is supported in part by the Simons Foundation Grant No. 488653. ALF is supported by the US Department of Energy Office of Science under Award Number DE-SC0015845, the Simons Collaboration on the Non-Perturbative Bootstrap, and a Sloan Foundation fellowship. AM is supported by the Simons Foundation Grant No. 385602 and the Natural Sciences and Engineering Research Council of Canada (NSERC), funding reference number SAPIN-2020-00047. EP is supported by the World Premier International Research Center Initiative,
MEXT, Japan, by the U.S. Department of Energy, Office of Science, Office of High Energy
Physics, under Award Number DE-SC0011632, and by ERC Starting Grant 853507.

\appendix

\section{Real analytic Eisenstein series and cusp forms}\label{app:eisensteinCusp}
\subsection{Real analytic Eisenstein series}\label{subApp:eisenstein}

The real analytic Eisenstein series are a special class of non-holomorphic modular functions formally defined in terms of a \emph{Poincar\'e series}, a sum over images of the modular group $PSL(2,\ZZ)$:
\begin{equation}\label{eq:eisensteinDefinition}
	E_s(\tau) = \sum_{\gamma\in\Gamma_\infty\backslash PSL(2,\ZZ)}\im(\gamma\tau)^s = y^s + \sum_{\substack{(c,d) = 1\\ c\ge 1}}{y^s\over |c\tau+d|^{2s}},\quad \re s > 1,
\end{equation}
where $\Gamma_\infty$ is the subgroup of $PSL(2,\ZZ)$ of  upper triangular matrices $\begin{pmatrix}1 & n \\ 0 & 1\end{pmatrix},\, n\in \ZZ,$ that fixes $y\coloneqq\im \tau$. The sum in (\ref{eq:eisensteinDefinition}) only converges for $\re s>1$, but it admits a meromorphic continuation to the entire $s$ plane, with the following Fourier decomposition
\begin{equation}\label{eq:eisensteinFourier}
	E_s(\tau) = y^{s} + {\Lambda(1-s)\over \Lambda(s)}y^{1-s} + \sum_{j=1}^\infty 4\cos(2\pi j x){\sigma_{2s-1}(j)\over j^{s-\half}\Lambda(s)}\sqrt{y}K_{s-\half}(2\pi j y),
\end{equation} 
where $\sigma_n(x) = \sum_{d|x}d^n$ is the divisor function and $K$ is the modified Bessel function of the second kind. The Fourier decomposition (\ref{eq:eisensteinFourier}) makes it clear that the Eisenstein series satisfies the following functional equation
\begin{equation}\label{eq:eisensteinFunctionalEqn}
	\Lambda(s)E_s(\tau) = \Lambda(1-s)E_{1-s}(\tau).
\end{equation}
The Eisenstein series are eigenfunctions of the Laplacian on the upper half-plane, with eigenvalue
\begin{equation}
	\Delta_\tau E_s(\tau) = s(1-s)E_s(\tau).
\end{equation}

The Eisenstein series are also eigenfunctions of the Hecke operators $T_j$
\begin{equation}
    T_j E_s(\tau) = {\sigma_{2s-1}(j)\over j^{s-\half}}E_s(\tau)
\end{equation}
where the action of the Hecke operators is defined by
\begin{equation}
\begin{aligned}\label{eq:heckeDefinition}
	T_j f(\tau) &= {1\over \sqrt{j}}\sum_{\substack{ad = j,\, d>0 \\ 0\le b \le d-1}}f\left({a\tau+b\over d}\right)\\
	&= {1\over \sqrt{j}}\left[\sum_{n=0}^{j-1}f\left({\tau+n\over j}\right)+f(j\tau)\right],~ j\text{ prime}.
\end{aligned}
\end{equation}
The Hecke operators satsify the multiplication formula
\begin{equation}
	T_j T_k = \sum_{\substack{\ell \, | \, (j,k)\\ \ell > 0}} T_{jk\over \ell^2}.
\end{equation}

The meromorphic continuation of the Eisenstein series has a simple pole at $s=1$ with a moduli-independent residue ${3\over \pi} = \text{vol}(\mathcal{F})^{-1}$. Moreover, the regular part of the Eisenstein series at $s=1$ is given by the Kronecker limit formula (see e.g. \cite{Terras_2013})
\begin{equation}
\begin{aligned}
	\widehat E_1(\tau) &\coloneqq \lim_{s\to 1}\left(E_s(\tau) - {3\over\pi(s-1)}\right)\\
    &= {6\over \pi}\left[1-12\zeta'(-1)-\log(4\pi\sqrt{y}|\eta(\tau)|^2)\right]\\
    &= y - \frac3\pi \log y+ {6\over \pi}\left(1-12\zeta'(-1)-\log(4\pi)\right) + {12\over \pi}\sum_{j=1}^\infty \frac{\cos(2\pi j x)\sigma_{1}(j)e^{-2\pi j y}}j.
\end{aligned}
\label{eq:e1hatdef}
\end{equation}
The regular part is not an eigenfunction of the Laplacian; rather, it satisfies the inhomogeneous equation
\begin{equation}
	\Delta_\tau \widehat E_1(\tau) = -{3\over \pi}.
\end{equation}

\subsection{Maass cusp forms}\label{subApp:cusp}

The Maass cusp forms are another important family of modular-invariant eigenfunctions of the Laplacian $\Delta_\tau$
\begin{equation}
	\Delta_\tau \nu_n(\tau) = \left({1\over 4}+R_n^2\right)\nu_n(\tau),
\end{equation} 
where $R_n$ is a positive real number. The cusp forms are either even or odd under parity, which we denote as $\nu^+_n(\tau)$ and $\nu^-_n(\tau)$ respectively, namely
\begin{align}
	\nu^+_n(\tau) &= \nu^+_n(-\bar \tau) \nonumber \\
	\nu^-_n(\tau) &= -\nu^-_n(-\bar\tau).
\end{align}
When we write $\nu_n(\tau)$ without specifying the superscript, we refer to both even and odd cusp forms. The cusp forms decay exponentially at the cusp $y=\infty$ and admit the following Fourier decompositions:
\begin{align}
	\nu^+_n(\tau) &= \sum_{j=1}^\infty a_j^{(n,+)}\cos(2\pi j x)\sqrt{y} K_{iR^+_n}(2\pi j y) \nonumber\\
	\nu^-_n(\tau) &= \sum_{j=1}^\infty a_j^{(n,-)}\sin(2\pi j x)\sqrt{y} K_{iR^-_n}(2\pi j y).
	\label{eq:nuexpansionasdf}
\end{align}
In the math literature, it is often convenient to adopt conventions such that the cusp forms are \emph{not} unit-normalized, in particular with the spin-one Fourier coefficient set to unity
\begin{equation}\label{eq:funnyNormalization}
	a_1^{(n,\pm)} = 1.
\end{equation}
We have listed the eigenvalues and low-lying Fourier coefficients of the cusp forms with smallest eigenvalues in Table \ref{table:1}.

\begin{table}[h!]
\centering
\begin{tabular}{| c c c c c c c c |} 
 \hline
 $n$ & $R^+_n$ & $\lambda^+_n = (R^+_n)^2 + \frac14$ & $(\nu^+_n, \nu^+_n)$ & $a_1^{(n,+)}$ & $a_2^{(n,+)}$ & $a_3^{(n,+)}$ & $a_4^{(n,+)}$  \\ 
 \hline
 1 & 13.780 & 190.132 & $4.539 \times 10^{-20}$ & $1$ & $1.549$ & $0.247$ & $1.400$ \\ 
 2 & 17.739 & 314.907 & $1.364 \times 10^{-25}$ & $1$&$-0.765$ & $-0.978$ & $-0.414$ \\
 3 & 19.423 & 377.522 & $8.167 \times 10^{-28}$ & $1$& $-0.693$ & $1.562$ & $-0.520$\\
 4 & 21.316  & 454.613 & $2.814 \times 10^{-30}$ & $1$&$1.288$ & $1.252$ & $0.658$\\
 5 & 22.786 & 519.448 & $1.380 \times 10^{-32}$ & $1$&$0.268$ & $-0.585$ & $-0.928$\\ 
 \hline \hline
  $n$ & $R^-_n$ & $\lambda^-_n = (R^-_n)^2 + \frac14$ & $(\nu^-_n, \nu^-_n)$ & $a_1^{(n,-)}$ & $a_2^{(n,-)}$ & $a_3^{(n,-)}$ & $a_4^{(n,-)}$ \\
  \hline 
  1 & 9.534 & 91.141 & $1.679 \times 10^{-14}$ & $1$ & $-1.068$ & $-0.456$ & $0.141$ \\
  2 & 12.173 & 148.432 & $4.239 \times 10^{-18}$ & $1$ & $0.289$ & $-1.202$ & $-0.916$ \\
  3 & 14.359 & 206.417 & $4.364 \times 10^{-21}$ & $1$ & $-0.231$ & $0.696$ & $-0.947$ \\
  4 & 16.138 & 260.687 & $2.622 \times 10^{-23}$ & $1$ & $1.162$ & $-1.282$ & $0.350$ \\
  5 & 16.644 & 277.281 & $6.636 \times 10^{-24}$ & $1$ & $-1.540$ & $0.977$ & $1.372$ \\
  \hline
\end{tabular}
\caption{Some numerical data for the first five even cusp forms, $\nu^+_n(\tau)$, and the first five odd cusp forms, $\nu^-_n(\tau)$. For more numerical precision, see \cite{LMFDB}. For much more numerical precision, see \cite{Booker}. 
}
\label{table:1}
\end{table}

Under this normalization, the Fourier coefficients $a_j^{(n)}$ are the eigenvalues of the cusp forms under the action of the Hecke operators $T_j$ defined in (\ref{eq:heckeDefinition}) \cite{maass1983lectures}
\begin{equation}
	T_j\nu_n(\tau) = a_j^{(n)}\nu_n(\tau).
	\label{eq:ajeigenhecke}
\end{equation}
The Fourier coefficients thus satisfy the Hecke relations
\begin{equation}\label{eq:HeckeRelation}
	a^{(n)}_j a^{(n)}_k = \sum_{\substack{\ell \, | \, (j,k)\\ \ell > 0}} a^{(n)}_{jk\o \ell^2}~.
\end{equation}
One simple application of this is found by setting $k=1$, so that 
\be
a^{(n)}_j a^{(n)}_1 = a^{(n)}_j 
\ee
From this we see that $a^{(n)}_1$ cannot vanish, or else all of the other coefficients (and hence the cusp form $\nu_n(\tau)$ itself) must vanish.  Moreover it sets $a^{(n)}_1=1$.
More generally, equation (\ref{eq:HeckeRelation}) implies that the only independent data in the cusp forms are the Fourier coefficients $\{a^{(n)}_j\}$ for prime spin $j$, as all others can be obtained as polynomials of the prime coefficients. For example from (\ref{eq:HeckeRelation}), we see that $a_4^{(n)} = (a_2^{(n)})^2-1$, as can be readily checked in Table \ref{table:1}. See section \ref{subsec:chaosAndCusps} for more discussion of properties satisfied by the Fourier coefficients of the cusp forms.

\section{Resolvent for the \texorpdfstring{$c=1$}{c1} free boson}\label{app:c=1Resolvent}

Here we will employ the spectral decomposition (\ref{eq:c=1Spectral}) of the primary partition functions of the $c=1$ free boson to compute the corresponding resolvent. 

The resolvent $C_j(\Delta)$ is a quantity that we define to have simple poles at the spectrum $\mathcal{I}_j$ of local primary operators of spin $j$, with residue given by the degeneracy $d_{\Delta_i,j}$ of the associated primary 
\begin{equation}\label{resolv}
	C_{j}(\Delta) \supset \sum_{\Delta_i\in\mathcal{I}_j}{d_{\Delta_i,j}\over \Delta-\Delta_i}.
\end{equation}
In 2d CFT with $c>c_{\rm currents}$, the resolvent \label{resolv} is not a well-defined quantity when applied to the full primary partition function $Z_p(\t)$, due to the lack of square-integrability from low-dimension operators, and the exponential growth of high dimension operators implied by the Cardy formula. However, $Z_{\rm spec}(\t)$ does admit a resolvent, as does $Z_p(\t)$ in certain special cases; the latter includes the Narain theories at $c=1$, where the analog of the Cardy asymptotic density of primaries behaves like (e.g. \cite{Afkhami-Jeddi:2020ezh, Maloney:2020nni})
\begin{equation}
	\rho^{(c=1)}_j(\Delta \gg 1) \sim \left(\Delta^2-j^2\right)^{-\half}.
\end{equation}
In such situations, where $\rho_j(\Delta\gg1)$ decays and hence the resolvent is well-defined, we may define the corresponding object, call it $Z(\t)$, in terms of the resolvent as 
\begin{equation}
	Z(\tau) = - y^{1\o 2}\sum_{j\in\ZZ}e^{2\pi i x j}\int_{\gamma}{d\Delta\over 2\pi i}\, C_{j}(\Delta)e^{-2\pi\Delta y},
\end{equation}
where $\gamma$ is a suitable vertical contour in the $\Delta$ plane. We thus extract the resolvent from a Laplace transform of the spin-$j$ partition function
\begin{equation}
	C_{j}(u) = -2\pi \int_0^\infty dy\, y^{-\half}e^{2\pi u y}Z_{j}(y),
\end{equation}
where for the purposes of computing the integral we assume $\re u < |j|$.

\subsection*{Scalar sector}
In what follows it will be convenient to define the general quantity
\begin{equation}
	H(t) \coloneqq {\left(Z,E_{\half+i t}\right)\over \Lambda\left(\half+i t\right)} 
\end{equation}
so that the contribution of the Eisenstein series in the spectral decomposition to the resolvent is given by
\begin{equation}
	C_0^{(E)}(u) = -\half \int_0^\infty dy \,e^{2\pi u y}\int_{-\infty}^\infty dt\, H(t)\left(\Lambda(-it)y^{it}+\Lambda(it)y^{-it}\right).
\end{equation}
In order to exchange the orders of integration we just need to assume $\re u < 0$. Then we have
\begin{equation}\label{eq:spin0resolvent}
	C^{(E)}_0(u) = -\half \int_{-\infty}^\infty dt\, H(t)\left[(-2\pi u)^{-1-i t}\pi^{it}\Gamma(1+it)\Gamma(-it)\zeta(-2it)+(t\to-t)\right].
\end{equation}
We now specify to the $c=1$ free boson, $Z(\t)=Z_p^{(c=1)}(\t)$, whereupon
\begin{equation}
H(t) =  4 \cos(t L_r),\quad L_r \coloneqq \log r^2
\end{equation}
In order to reproduce the lattice of poles corresponding to the scalar spectrum of the free boson theory, we would like to write the zeta function above as an infinite sum $\zeta(-2it) = \sum_{n=1}^\infty n^{2it}$. In order to do this, we need to shift the $t$ contour so that $\im t >\half$ for the first term above, and $\im t < -\half$ for the second term. Focusing on the first term, in doing so we pick up the residue of the integrand at $t = {i\over 2}$, in addition to a contribution from a pole that develops at $t = 0$ due to the fact that we are deforming the contour for the two terms in different directions, which can be dealt with via a principal value prescription. Referring to the new $t$ contour as $\mathcal{C}$, we then have
\begin{equation}
\begin{aligned}\label{eq:spin0resolvent2}
	C^{(E)}_0(u) &= {\sqrt{2}\pi(r+r^{-1})\over \sqrt{-u}}+{1\over u} -\sum_{n=1}^\infty\int_\mathcal{C} dt\, H(t)(-2\pi u)^{-1-i t}\pi^{i t}\Gamma(1+i t)\Gamma(-it)n^{2it}\\
	&={\sqrt{2}\pi(r+r^{-1})\over \sqrt{-u}} +{1\over u}+ 2\sum_{n=1}^\infty \sum_{m=1}^\infty\left(-{2\over u}\right)\left({2u\over n^2}\right)^m \cosh(m L_r)\\
	&={\sqrt{2}\pi(r+r^{-1})\over \sqrt{-u}} +{1\over u}+  2\sum_{n=1}^\infty \left({1\over u-n^2 r^2/2}+{1\over u-n^2 r^{-2}/2}\right),
\end{aligned}
\end{equation}
where we have evaluated the integral in the first line by deforming the contour upwards in the complex $t$ plane and summing over residues at $t = i\mathbb{Z}_{>0}$.

Meanwhile, the contribution of the constant term to the resolvent is given by
\begin{equation}\label{eq:cequalsonec0}
	C_0^{(0)}(u) = -2\pi \int_0^\infty dy\, y^{-1/2}e^{2\pi u y}(r+r^{-1}) = -{\sqrt{2}\pi (r+r^{-1})\over\sqrt{-u}},
\end{equation}
exactly canceling the first term in (\ref{eq:spin0resolvent2}).

\subsection*{Spinning sectors}
So far we have seen that the spectral decomposition (\ref{eq:c=1Spectral}) of the $c=1$ free boson leads to poles in the scalar resolvent at precisely the locations of local scalar operators. This is not unexpected since the Maass cusp forms have no scalar components. Here we will see how the correct spectrum is generated in the spinning sectors. The contribution to the spinning resolvent from the Eisenstein series is given by
\begin{equation}
	C_{j}^{(E)}(u) = -\int_{0}^\infty dy\, e^{2\pi u y}\int_{-\infty}^\infty dt\, H(t){2\sigma_{2it}(j)\over j^{it}}K_{it}(2\pi j y)
\end{equation}
Again for the meantime assuming $\re(u)< j$, we can exchange the integrals to arrive at 
\begin{equation}
	C_{j}^{(E)}(u) = -{1\over \sqrt{u^2-j^2}}\int_{-\infty}^\infty dt\,H(t){\sigma_{2it}(j)\over j^{it}}{\sin (tL_u)\over \sinh(\pi t)},
\end{equation}
where we have defined
\begin{equation}
	L_u \coloneqq \log\left(-{j\over u+\sqrt{u^2-j^2}}\right).
\end{equation}
Writing $H(t){\sigma_{2it}(j)j^{-it}} = 4\sum_{d|j}\cos(tL_r)\cos(t L_d)$, where $L_d \coloneqq \log\left({j\over d^2}\right)$, we then have
\begin{equation}
\begin{aligned}
	C_{j}^{(E)}(u) &= -{4\over\sqrt{u^2-j^2}}\sum_{d|j}\int_{-\infty}^\infty dt\,{\cos(tL_r)\cos(tL_d)\sin(t L_u)\over\sinh(\pi t)}\\
	&= -{4\over \sqrt{u^2-j^2}}\sum_{d|j}1 + {8\over \sqrt{u^2-j^2}}\sum_{d|j}\sum_{n=1}^\infty(-)^{n+1}e^{-nL_u}\cosh(nL_d)\cosh(nL_r)\\
	&=-{4\over \sqrt{u^2-j^2}}\sum_{d|j}1 +{2\over \sqrt{u^2-j^2}}\sum_{d|j}\left(2+{\sqrt{u^2-j^2}\over u-({j^2\over 2d^2 r^2}+{d^2 r^2\over 2})}+{\sqrt{u^2-j^2}\over u - ({d^2\over 2r^2}+{j^2 r^2\over 2 d^2})}\right)\\
	&= \sum_{d|j}\left({2\over u-({j^2\over 2d^2 r^2}+{d^2 r^2\over 2})}+{2\over u -({d^2\over 2r^2}+{j^2 r^2\over 2 d^2})}\right).
\end{aligned}
\end{equation}
In the second line we wrote the sine as a difference of complex exponentials, deformed the $t$ contour in appropriate directions for each term, picked up an appropriate residue from a pole at $t=0$ according to a principal value prescription (this is the first term), and summed over the residues of the poles of $\sinh(\pi t)^{-1}$. What we are left with is precisely the discrete sum over poles corresponding to the states of spin $j$ in the $c=1$ free boson at radius $r$. So the cusp forms must play no role in reconstructing the discrete sum over operators in the $c=1$ partition functions. Indeed in appendix \ref{app:derivingEight} we will prove explicitly that the overlap of the partition function of the $c=1$ free boson with the cusp forms vanishes.

\section{Derivation of cusp form overlap for $c=1$ and $c=2$ Narain theories}\label{app:derivingEight}
In this appendix we will derive the inner product between the primary partition function of the $c=1$ (\ref{eq:c=1}) and $c=2$ (\ref{eq:c=2Narain}) Narain theories  and the Maass cusp forms\footnote{We are grateful to Herman Verlinde for discussions that inspired the computation in this appendix.}
\begin{eqnarray}
    \left(Z^{(c=1)},\nu_n\right) &=& 0, \\
        \left(Z^{(c=2)},\nu^+_n\right) &=& 8\nu^+_n(\rho)\nu^+_n(\sigma), \\
            \left(Z^{(c=2)},\nu^-_n\right) &=& -8i\nu^-_n(\rho)\nu^-_n(\sigma), 
\end{eqnarray}
where $\rho$ and $\sigma$ are respectively the complexified Kahler structure and the complex structure of the target $T^2$. In general CFTs, we do not know how to compute the inner product with the cusp forms except by brute-force numerical integration because one does not have access to the unfolding trick. However, if the partition function can be rewritten as a sum over terms that are themselves Poincar\'e series, then one can apply the unfolding trick to each of those terms.  We will show how to do this explicitly for the $c=1$ and $c=2$ Narain lattices.  The starting point is to note that the Narain lattice partition function after Poisson resummation, equation (\ref{eq:PoissonNarain}),  is manifestly modular invariant, and moreover that an $SL(2,\mathbb{Z})$ transformation $\tau \rightarrow \frac{a \tau + b}{c \tau + d}$ on the complex modulus $\tau$ can be absorbed by a transformation $(m^i, n^i) \rightarrow (d m^i + b n^i, c m^i+a n^i)$ on the set of integers $m^i$ and $n^i$ in the sum.  This provides an action of $SL(2,\mathbb{Z})$ on the $(m^i, n^i)$s, which lets us group them into irreducible representations.  We can then write the total sum over $(m^i, n^i)$ as a sum over these irreducible representations, i.e. as a sum over Poincar\'e series. 

We begin with the simpler case of the $c=1$ Narain lattice.  In this case, every pair $(m',n')$ can be obtained by a $\Gamma_\infty \backslash PSL(2,\mathbb{Z})$ transformation acting on $(m,0)$ where $m$ is the greatest common factor of $(m',n')$.  This fact follows immediately from the general $SL(2,\mathbb{Z})$ transformation acting on $(m,0) \rightarrow (dm, cm)$.  Therefore, we can compute the overlap of $Z^{(c=1)}$ with a cusp form $\nu_n$ as follows:
\begin{equation}
\Big( Z^{(c=1)}, \nu_n \Big) = \int_{\mathcal{F}} \frac{d^2 \tau}{y^2} Z^{(c=1)}(\tau; r) \nu_n(\tau)=  2 r \sum_{m=1}^\infty \int_{\Gamma_\infty \backslash \mathbb{H}} \frac{d^2 \tau}{y^2} e^{ - \pi \frac{m^2 r^2}{y}} \nu_n(\tau) =0,
\end{equation}
where we have used the fact that $\nu_n$ has no scalar part.

Next, to write the $c=2$ Narain partition function in terms of Poincar\'e series, it is helpful to  combine the momentum and winding numbers $n_1,n_2,w^1,w^2$ into a matrix whose determinant is (minus) the spin of the corresponding primary operator in the sum defining the lattice theta function. In the case that the determinant is unity, this corresponds to a Poincar\'e series.\footnote{A similar manipulation involving the unfolding trick was used in \cite{Callebaut:2019omt} in developing the worldsheet formulation of the $T\bar T$ deformation of 2d CFTs. There, the analog of (\ref{eq:Z1}) was the ``wrapping number one'' sector of the part of the worldsheet CFT with $T^2$ target.} To move to sectors of higher spin, one applies a generalization of the Hecke operators. Concretely, (\ref{eq:c=2Narain}) can be rewritten as
\begin{equation}
\begin{aligned}\label{eq:c=2New}
    Z^{(c=2)}(\tau;\rho,\sigma) =& \, y + 2y\sum_{n=1}^\infty\sum_{\gamma_1,\gamma_2\in\Gamma_\infty\backslash PSL(2,\mathbb{Z})}\exp\left[-{\pi n^2 y\over \im(\gamma_1\sigma)\im(\gamma_2\rho)}\right]\\
    & \, + \sum_{j=1}^\infty \widetilde T_j\left[Z_{-1}^{(c=2)}(\tau;\rho,\sigma) + Z_{1}^{(c=2)}(\tau;\rho,\sigma)\right],
\end{aligned}
\end{equation}
where $Z_{-1}^{(c=2)}$ and $Z_{1}^{(c=2)}$ are Poincar\'e series
\begin{equation}\label{eq:Z1}
    Z_{\mp 1}^{(c=2)}(\tau;\rho,\sigma) = 2y\sum_{\gamma\in PSL(2,\mathbb{Z})}\exp\left[{\pi i \over 2}\left(\tau{|\rho\mp\gamma\sigma|^2\over\im(\rho)\im(\gamma\sigma)} + {\bar\tau}{|\rho\mp\gamma\bar\sigma|^2\over\im(\rho)\im(\gamma\bar\sigma)}\right)\right]
\end{equation}
and $\widetilde T_j$ is a generalization of the Hecke operator (\ref{eq:heckeDefinition}) that acts on $\tau$ as well as $\sigma$
\begin{equation}
    \widetilde T_N Z(\tau;\rho,\sigma) = {1\over N}\sum_{\substack{ad=N,\, d>0 \\ 0 \le b\le d-1}}Z\left(N\tau;\rho,{a\sigma+b\over d}\right).
\end{equation}
Importantly, these Hecke operators give the same answer acting on either $\rho$ or $\sigma$, so we could also write
\begin{equation}
    \widetilde T_N Z(\tau;\rho,\sigma) = {1\over N}\sum_{\substack{ad = N,\, d>0 \\ 0 \le b \le d-1}}Z\left(N\tau;{a\rho+b\over d},\sigma\right).
\end{equation}
In what follows it will be important that the partition function (\ref{eq:c=2New}) is invariant under the triality exchanging any pair of $\tau,\rho,\sigma$, provided the action of the Hecke operators $\widetilde T$ is suitably modified.

To get started, we consider the inner product between an even cusp form $\nu^+_n$ and the Poincar\'e series $Z_{\mp 1}$ as defined in (\ref{eq:Z1}) (up to a change of triality frame)
\begin{equation}\label{eq:Z1InnerProduct}
\begin{aligned}
    \left(Z_{\mp 1}^{(c=2)},\nu^+_n\right) &= \int_{\mathcal{F}}{d^2\tau\over y^2}\, Z_{\mp 1}^{(c=2)}(\rho;\sigma,\tau)\nu^+_n(\tau)\\
    &= 2\rho_2 \int_{\mathbb{H}}{d^2\tau\over y^2}\, e^{{\pi i \over 2}\left({\rho\over \sigma_2 y}|\sigma\mp\tau|^2 - {\bar\rho\over \sigma_2 y}|\sigma\mp\bar\tau|^2\right)}\nu^+_n(\tau)\\
    &= 2\sum_{j=1}^\infty a_j^{(n)}\sqrt{\rho_2\sigma_2}e^{\mp 2\pi i \rho_1}\cos(2\pi j \sigma_1)\int_0^\infty {dy\over y}\,e^{-\pi\rho_2\left({\sigma_2\over y}+{y\over \sigma_2}\right)-\pi j^2{\sigma_2 y\over \rho_2}}K_{iR^+_n}(2\pi j y),
\end{aligned}
\end{equation}
where in the second line we unfolded the integral over the fundamental domain to one over the upper half-plane, and in the third we used the explicit form of the Fourier decomposition of the Maass cusp forms (\ref{eq:nuexpansionasdf}). The remaining integral over the imaginary part of $\tau$ is given by
\begin{equation}
    \int_0^\infty {dy\over y}\, e^{-\pi\rho_2\left({\sigma_2\over y}+{y\over \sigma_2}\right)-\pi j^2{\sigma_2 y\over \rho_2}}K_{iR^+_n}(2\pi j y) = 2 K_{iR^+_n}(2\pi\rho_2)K_{iR^+_n}(2\pi j \sigma_2)
\end{equation}
so that the inner product (\ref{eq:Z1InnerProduct}) assembles into
\begin{equation}
     \left(Z_{\mp 1}^{(c=2)},\nu^+_n\right) = 4\sqrt{\rho_2}e^{\mp 2\pi i \rho_1}K_{iR^+_n}(2\pi \rho_2)\nu^+_n(\sigma).
\end{equation}

To derive the overlap with the remaining parts of the $c=2$ partition function, we simply apply the Hecke operators $\widetilde T_N$, which now rescale $\rho\to N\rho$ and sum over $\sigma\to {a\sigma+b\over d}$:\footnote{We would get the same answer if we took the Hecke operators to rescale $\rho\to N\rho$ and sum over $\tau\to {a\tau+b\over d}$. In this case we would be computing $\sum_{N=1}^\infty \left[\left(\widetilde T_N Z_{-1}^{(c=2)},\nu^+_n\right) + \left(\widetilde T_N Z_{1}^{(c=2)},\nu^+_n\right)\right]$, which reduces to (\ref{eq:c=2HeckeSum}) upon recognizing that the ordinary Hecke operators $T_N$ are self-adjoint with respect to the Petersson inner product.} 
\begin{equation}
\begin{aligned}\label{eq:c=2HeckeSum}
    \left(Z^{(c=2)},\nu^+_n\right) &= \sum_{N=1}^\infty \widetilde T_N\left[\left( Z^{(c=2)}_{-1},\nu^+_n\right)+\left( Z^{(c=2)}_{1},\nu^+_n\right)\right]\\
    &= 4\sqrt{\rho_2} \sum_{N=1}^\infty \left(e^{-2\pi i N \rho_1} + e^{2\pi i N\rho_1}\right)K_{iR^+_n}(2\pi N \rho_2) T_N\nu^+_n(\sigma)\\
    &= 8 \sqrt{\rho_2}\sum_{N=1}^\infty \cos(2\pi N \rho_1)a_N^{(n,+)}K_{iR^+_n}(2\pi N \rho_2)\nu^+_n(\sigma)\\
    &= 8\nu^+_n(\rho)\nu^+_n(\sigma).
\end{aligned}
\end{equation}
where $T_N$ is the conventional Hecke operator defined in (\ref{eq:heckeDefinition}), and we have made use of the fact that the cusp forms are eigenfunctions of the Hecke operators as in (\ref{eq:ajeigenhecke}). Note that the difference in normalization between $\widetilde T_N$ and $T_N$ was essential for this to work out.

An essentially identical calculation shows that
\be
 \left(Z^{(c=2)},\nu^-_n\right) = -8i\nu^-_n(\rho)\nu^-_n(\sigma).
\ee

\section{Clebsch-Gordan coefficients for Eisenstein series and cusp forms}\label{app:eisensteinCG}

It is often interesting to consider partition functions in the product locus of the Narain moduli space, namely, partition functions in $\mathcal{M}_{c_1+c_2}$ that can be written as a product of partition functions in $\mathcal{M}_{c_1}$ and $\mathcal{M}_{c_2}$. For these purposes, one must understand the spectral decomposition of products of eigenfunctions of the worldsheet Laplacian. We are grateful to Peter Humphries for explaining the results of Appendix \ref{sec:computingcg} to us.

\subsection{Computing the Clebsch-Gordan coefficients}
\label{sec:computingcg}

We will begin by considering the inner product of a product of real analytic Eisenstein series with another Eisenstein series. This inner product requires regularization, so we will write it as a RS transform, which can be computed using the methods of Section \ref{subSec:RankinSelberg}
\begin{equation}
	K_{s_1,s_2,s_3} \coloneqq 	R_{s_1}[E_{s_2}E_{s_3}].
\end{equation}
Using the RS method as adapted by Zagier (indeed, this example was treated in Section 3 of \cite{zbMATH03796039}), after discarding the terms of slow growth at the cusp in the scalar sector of $E_{s_2}E_{s_3}$, this inner product can be written in terms of a Mellin transform 
\begin{equation}
\begin{aligned}
	K_{s_1,s_2,s_3} =& \,  {8\over \Lambda(s_2)\Lambda(s_3)} \int_0^\infty dy \, y^{s_1-2}\sum_{j=1}^\infty j^{1-s_2-s_3}\sigma_{2s_2-1}(j)\sigma_{2s_3-1}(j)\, y\, K_{s_2-\half}(2\pi j y)K_{s_3-\half}(2\pi j y)
\end{aligned}
\end{equation}
By making appropriate assumptions about the real parts of the $s_i$, we can exchange the sum and the integral to arrive at
\begin{equation}
	K_{s_1,s_2,s_3} = {1\over \Lambda(s_1)\Lambda(s_2)\Lambda(s_3)}\prod_{\pm_{1,2}} \Lambda\left({s_1\pm_1\left(s_2-\half\right)\pm_2\left(s_3-\half\right)\over 2}\right).
\end{equation}
This result is completely symmetric under exchange of $s_1,s_2,s_3$ due to the functional equation satisfied by $\Lambda(s)$.

Next we will consider the overlap of two real analytic Eisenstein series with an even Maass cusp form\footnote{The overlap of two Eisenstein series with an odd Maass cusp form vanishes.}
\begin{equation}
	K^n_{s_1,s_2} \coloneqq \int_{\mathcal{F}}{dxdy\over y^2}\, \nu^+_n(\tau)E_{s_1}(\tau)E_{s_2}(\tau).
\end{equation}
The presence of the cusp form means that this does not require regularization in the same sense as the previous case of three Eisenstein series. This can also be written as a Mellin transform
\begin{equation}
	K^n_{s_1s_2} = \int_0^\infty dy \, y^{s_1-1}\sum_{j=1}^\infty \left({2\sigma_{2s_2-1}(j)\over j^{s_2-\half}\Lambda(s_2)}K_{s_2-\half}(2\pi j y)\right)a_j^{(n,+)} K_{iR^+_n}(2\pi j y).
\end{equation}
Given appropriate assumptions about the real parts of $s_1,s_2$, we can exchange the sum and the integral to arrive at
\begin{equation}
	K^n_{s_1s_2} = {\pi^{-2s_1}\zeta(2s_1)\over 4\Lambda(s_1)\Lambda(s_2)}\left[\sum_{j=1}^\infty a_j^{(n,+)}j^{\half-s_1-s_2}\sigma_{2s_2-1}(j)\right]\prod_{\pm_{1,2}}\Gamma\left({s_1\pm_1\left(s_2-\half\right)\pm_2 i R^+_n\over 2}\right).
\end{equation}
The sum over $j$ can be simplified by making use of the fact that the cusp form Fourier coefficients are eigenvalues of the Hecke operators (c.f. (\ref{eq:HeckeRelation})) to obtain
\begin{equation}
	\zeta(2s_1)\sum_{j=1}^\infty {a_j^{(n,+)}\over j^{s_1}}\sum_{ab=j}\left({a\over b}\right)^{s_2-\half} = \sum_{j=1}^\infty {a_j^{(n,+)}\over j^{s_1+s_2-\half}}\sum_{\ell=1}^\infty{a_{\ell}^{(n,+)}\over \ell^{s_1-s_2+\half}} = L^{(n)}\left(s_1+s_2-\half\right)L^{(n)}\left(s_1-s_2+\half\right),
\end{equation}
where $L^{(n)}$ is a so-called \textbf{$L$-function}, a Dirichlet series involving the Fourier coefficients of cusp forms 
\begin{equation}
	L^{(n)}(s) \coloneqq \sum_{j=1}^\infty {a_j^{(n,+)}\over j^s}.
\end{equation}
Although the above expression only converges for $\re s>1$, the $L$-functions admit a meromorphic continuation to the entire $L$ plane and satisfy a functional equation. In particular, defining
\begin{equation}
	\widetilde L^{(n)}(s) \coloneqq \pi^{-s}\Gamma\left(\half\left(s+iR^+_n\right)\right)\Gamma\left(\half\left(s-iR^+_n\right)\right)L^{(n)}(s)
\end{equation}
we have
\begin{equation}
	\widetilde L^{(n)}(s) = \widetilde L^{(n)}(1-s).
\end{equation}
and is entire. All together, we then have
\begin{equation}
	K^n_{s_1,s_2} = {1\over 4\Lambda(s_1)\Lambda(s_2)}\widetilde L^{(n)}\left(s_1+s_2-\half\right)\widetilde L^{(n)}\left(s_1-s_2+\half\right)
\end{equation}

We will now compute the final ``Clebsch-Gordan'' coefficient that involves an Eisenstein series and thus can be evaluated used the RS unfolding method, namely the overlap of two Maass cusp forms with an Eisenstein series\footnote{This is only nontrivial if the Maass forms have the same parity, so we proceed with the understanding that $K^{n_1,n_2}_s$ is only non-vanishing if $\nu_{n_1}$ and $\nu_{n_2}$ have the same parity.}
\begin{equation}
\begin{aligned}
	K^{n_1,n_2}_s &\coloneqq \int_{\mathcal{F}}{dxdy\over y^2} \, \nu^\pm_{n_1}(\tau)\nu^\pm_{n_2}(\tau)E_s(\tau)\\
	&= \half\int_0^\infty dy\, y^{s-1}\sum_{j=1}^\infty a_j^{(n_1,\pm)}a_j^{(n_2,\pm)}K_{iR^\pm_{n_1}}(2\pi j y)K_{iR^\pm_{n_2}}(2\pi j y).
\end{aligned}
\end{equation}
Exchanging the sum and integral to evaluate the Mellin transform as usual, we arrive at
\begin{equation}
\begin{aligned}
	K^{n_1,n_2}_s &= {\pi^{-2s}\zeta(2s)\over 16\Lambda(s)}\left[\sum_{j=1}^\infty {a_j^{(n_1,\pm)}a_j^{(n_2,\pm)}\over j^s}\right]\prod_{\pm_{1,2}}\Gamma\left({s\pm_1 iR^\pm_{n_1} \pm_2 i R^\pm_{n_2}\over 2}\right)\\
	&= {\pi^{-2s}\over 16\Lambda(s)}L^{(n_1,n_2)}(s) \prod_{\pm_{1,2}}\Gamma\left({s\pm_1 iR^\pm_{n_1} \pm_2 i R^\pm_{n_2}\over 2}\right),
\end{aligned}
\end{equation}
where we have recognized the sum over $j$ as another $L$-function
\begin{equation}
	L^{(n_1,n_2)}(s) \coloneqq \zeta(2s)\sum_{j=1}^\infty{a^{(n_1,\pm)}_j a^{(n_2,\pm)}_j\over j^s}. 
\end{equation}
Similarly to the other $L$-functions we've considered, this $L$-function admits a meromorphic continuation to the entire $s$ plane and satisfies the functional equation
\begin{equation}
	\widetilde L^{(n_1,n_2)}(s) = \widetilde L^{(n_1,n_2)}(1-s),
\end{equation}
where
\begin{equation}
	\widetilde L^{(n_1,n_2)}(s) \coloneqq \pi^{-2s}\left[\prod_{\pm_{1,2}}\Gamma\left({s\pm_1 iR^\pm_{n_1} \pm_2 i R^\pm_{n_2}\over 2}\right)\right]L^{(n_1,n_2)}(s).
\end{equation}
The residues of these Clebsch-Gordan coefficients at $s=1$ encode the norms of the cusp forms via
\begin{equation}
    (\nu^\pm_n,\nu^\pm_n) = {\pi\over 3}\Res_{s=1}K^{n,n}_s.
\end{equation}

Finally, there is the overlap coefficient involving three Maass cusp forms
\begin{equation}
	K^{n_1,n_2,n_3} \coloneqq \int_{\mathcal{F}}{dxdy\over y^2}\, \nu_{n_1}(\tau)\nu_{n_2}(\tau)\nu_{n_3}(\tau).
\end{equation}
Although there is no direct way to apply a RS transform in this integral, in the math literature it is known that the square of the average of certain triple products of Maass cusp forms over the fundamental domain can be written in terms of a certain $L$-function, see for example \cite{Humphries_2018,Ichino_2008,Watson:2002uw}.

\subsection{Two routes to the product locus}
The fact that the overlap coefficient involving two Eisenstein series and one cusp form is generically non-vanishing means that even if certain partition functions have spectral decompositions without support on cusp forms, the product partition function will have support on cusp forms. Indeed we can see explicitly that this is true by considering the case of the product of $c=1$ free boson partition functions (\ref{eq:c=1Spectral}) and the $c=2$ Narain partition functions (\ref{eq:c=2Spectral}) on the product locus.\footnote{These statements apply to the primary partition functions, $Z_p^{(c)} \times Z_p^{(c)} = Z_p^{(2c)}$, since the chiral algebra enlarges to $U(1)^{2c}$ upon squaring.} Although the former does not have support on cusp forms, the latter does. Indeed, the equality of the product of partition functions given by (\ref{eq:c=1Spectral}) and (\ref{eq:c=2Spectral}) on the product locus implies some remarkably non-trivial integral identities. 

Our spectral decompositions of the $c=1$ (\ref{eq:c=1Spectral}) and $c=2$ (\ref{eq:c=2Spectral}) imply the following equation:
\begin{equation}
\begin{aligned}\label{eq:productMiracles}
	&\, \prod_{j=1}^2\left[r_j+r_j^{-1} + {1\over 4\pi i}\int_{\re s_j = \half}ds_j \, 2\Lambda(s_j)\left(r_j^{1-2s_j}+r_j^{2s_j-1}\right)E_{s_j}(\tau)\right]\\
	=& \, \alpha + \widehat{E}_1(ir_1r_2) + \widehat{E}_1(ir_2/r_1) + \widehat{E}_1(\tau)\\
	& \, + {1\over 4\pi i}\int_{\re s =\half}ds\, 2{\Lambda(s)^2\over \Lambda(1-s)}E_s(ir_1r_2)E_s(ir_2/r_1)E_s(\tau)\\
	& \, + 8\sum_{n=1}^\infty(\nu^+_n,\nu^+_n)^{-1}\nu^+_n(ir_1r_2)\nu^+_n(ir_2/r_1)\nu^+_n(\tau).
\end{aligned}
\end{equation}
(Since the product locus is invariant under parity, there are no odd cusp forms on the RHS of (\ref{eq:productMiracles}).)
We would like to use our results for the Clebsch-Gordan coefficients $K$ to unpack this equation. These Clebsch-Gordan coefficients can be understood as coefficients in the spectral decomposition of the product of eigenfunctions of the worldsheet Laplacian. For example, we can use spectral decomposition to write the product of Eisenstein series appearing on the left-hand side of (\ref{eq:productMiracles}) as
\begin{equation}
\begin{aligned}\label{eq:productOfEisensteins}
	E_{s_1}(\tau)E_{s_2}(\tau) =& \, E_{s_1+s_2}(\tau) + \frac{\Lambda(1-s_1)}{\Lambda(s_1)}E_{1-s_1+s_2}(\tau) + \frac{\Lambda(1-s_2)}{\Lambda(s_2)}E_{1-s_2+s_1}(\tau) \\
	& \,+ \frac{\Lambda(1-s_1)\Lambda(1-s_2)}{\Lambda(s_1)\Lambda(s_2)} E_{2-s_1-s_2}(\tau) + {1\over 4\pi i }\int_{\re s=\half}ds K_{1-s,s_1,s_2}E_s(\tau)
	 \\ & \,+ \sum_{n=1}^\infty K_{s_1,s_2}^n (\nu^+_n,\nu^+_n)^{-1}\nu^+_n(\tau).
\end{aligned}
\end{equation}
In (\ref{eq:productOfEisensteins}), we sum only over even cusp forms as usual, and it is important to have $\re s_1 = \re s_2 = \frac12$. (We can move off this locus by simply removing any of the first four terms in (\ref{eq:productOfEisensteins}) where the Eisenstein series $E_s(\tau)$ has $\re s < \frac12$.)

Inspecting (\ref{eq:productMiracles}) and using the explicit form of $K_{s_1,s_2}^n$, we get the following identity by demanding the cusp forms of both sides match. For all $r_1, r_2$, and for all even cusp forms $\nu^+_n$, we claim:
\begin{align}
&\int_{\frac12-i\infty}^{\frac12+i\infty} ds_1 \int_{\frac12-i\infty}^{\frac12+i\infty} ds_2 (r_1^{2s_1-1} + r_1^{1-2s_1})(r_2^{2s_2-1} + r_2^{1-2s_2}) \pi^{-2s_1} \nn\\
&\times L^{(n)}\(s_1 + s_2 - \frac12\)L^{(n)}\(s_1 - s_2 + \frac12\) \prod_{\pm_{1,2}} \Gamma\(\frac{s_1 \pm_1 \(s_2-\frac12\) \pm_2 i R^+_n}2\) \nn\\
&= -128\pi^2 \nu^+_n(i r_1 r_2) \nu^+_n(i r_1/r_2). 
\end{align}  

\section{Some generalizations of Eisenstein series}\label{app:generalizedEisenstein}

In Section \ref{sec4}, we discussed ways to render general CFT partition functions in $L^2(\mathcal{F})$, by subtracting out modular completions of light operators. Recall that the operators that make a CFT primary partition function no longer square-integrable are those primaries which obey
\be
h + \bar{h} \leq \frac{c-c_{\rm currents}}{12}.
\label{eq:hhbc112}
\ee
For a compact CFT, there will always be a finite number of primary operators obeying (\ref{eq:hhbc112}). 

The Poincar\'e sum 
\be
\sum_{\gamma\in \Gamma_\infty \backslash PSL(2,\mathbb Z)} \(y^s e^{-2\pi y E} e^{2\pi i x J}\)\Big{|}_{\gamma}
\label{eq:poincaremkw}
\ee
was considered in \cite{Maloney:2007ud, Keller:2014xba}. Following these references, we will rewrite (\ref{eq:poincaremkw}) as
\begin{align}
\sum_{\gamma\in \Gamma_\infty \backslash PSL(2,\mathbb Z)} \(e^{-2\pi y E} e^{2\pi i x J}\)\Big{|}_{\gamma} &=   \sum_{m=0}^{\infty} \frac{(-2\pi E)^m}{m!} \sum_{\gamma\in \Gamma_\infty \backslash PSL(2,\mathbb Z)} \(y^{m+s} e^{2\pi i x J}\)\Big|_{\gamma} \nn\\
&=  \sum_{m=0}^{\infty} \frac{(-2\pi E)^m}{m!} E(m+s, \tau, J),
\end{align}
where we define $E(s,\tau,J)$ as a generalization of the Eisenstein series:
\be
E(s, \tau, J) \coloneqq \sum_{\gamma\in \Gamma_\infty \backslash PSL(2,\mathbb Z)} \(y^s e^{2\pi i J x}\)\Big|_\gamma.
\label{eq:eisgendef}
\ee
The Fourier coefficients of (\ref{eq:eisgendef}) were computed in \cite{Keller:2014xba} and are given by\footnote{To clarify (\ref{eq:FourierEis}): if any of the terms in the infinite sums diverge, it is important to first evaluate $a$ and then take the limit in $s$, as opposed to vice versa.}:
\begin{align}
E(s, \tau, J) &= e^{2\pi i J x} y^s + \sum_{a=0}^\infty \frac{(-1)^a\pi^{2a+\frac12} \sigma_{4a+2s-1}(|J|) \Gamma\(a+s-\frac12\) y^{1-2a-s}}{|J|^{2a+2s-1}\Gamma(a+1)\Gamma(2a+s)\zeta(4a+2s)} \nn\\
&+ \sum_{j \in \mathbb Z, j \neq 0} e^{2\pi i x j}\sum_{a=0}^\infty \frac{2\mathcal{Z}_{j,J}(a+s) \pi^{a+s} |j|^{s-\frac12}J^a\text{sgn}(j)^a}{\Gamma(a+1)\Gamma(a+s)} y^{1-a-s} \(\partial_{y}\)^a \(y^{a+s-\frac12}K_{a+s-\frac12}\(2\pi y |j|\)\)
\label{eq:FourierEis}
\end{align}
In (\ref{eq:FourierEis}), $\mathcal{Z}_{j,J}$ is a Kloosterman zeta function (see \cite{Keller:2014xba} for more detail). We have verified (\ref{eq:FourierEis}) to be modular invariant numerically. We can also simplify the derivatives of the Bessel functions to get:
\begin{align}
&E(s, \tau, J) = e^{2\pi i J x} y^s + \sum_{a=0}^\infty \frac{(-1)^a\pi^{2a+\frac12} \sigma_{4a+2s-1}(|J|) \Gamma\(a+s-\frac12\) y^{1-2a-s}}{|J|^{2a+2s-1}\Gamma(a+1)\Gamma(2a+s)\zeta(4a+2s)} \nn\\
&+ \sum_{j \in \mathbb Z, j \neq 0} e^{2\pi i x j}\sum_{a=0}^\infty \sum_{b=0}^{\lfloor \frac a2 \rfloor} \frac{(-1)^{a+b} 2^{1+a-2b}|j|^{a+s-\frac12}(2b+s-\frac12)\pi^{2a+s}\Gamma(s+b-\frac12)}{\Gamma(1+a-2b)\Gamma(b+1)\Gamma(2b+s+\frac12)\Gamma(a+s)} \nn\\
&~~~~~~~~~~~~~~~~~\times ~_2F_1\(\frac{2b-a}2, \frac{1-a+2b}2,2b+s+\frac12;1\) J^a \text{sgn}(j)^a\mathcal{Z}_{j,J}(a+s) y^{\frac12}K_{s+2b-\frac12}\(2\pi |j| y\)
\label{eq:FourierEisBetter}
\end{align}
As a check, (\ref{eq:FourierEisBetter}) reproduces Eqn (C.6) of \cite{Benjamin:2020mfz} in the special case of $E(s, \tau,\bar\tau, 1) + E(s, \tau,\bar\tau, -1)$.

If $s<\frac12$, we can perform a spectral decomposition of $E(s,\tau, J)$ by subtracting the finite number of terms in the first sum of (\ref{eq:FourierEis}), (\ref{eq:FourierEisBetter}) that scale as $y^{1/2}$ or faster at large $y$. For example, for $0<s<\frac12$, we get
\begin{align}
E(s,\tau,J) + E(s, \tau, -J) &= \frac{2\sigma_{1-2s}(|J|)}{\zeta(2s-1)} E_s(\tau)  \nn\\&+ \frac{1}{4\pi i}\int_{\frac12-i\infty}^{\frac12+i\infty} ds' \frac{|J|^{1-s-s'}\pi^{s'-s} \sigma_{2s'-1}(|J|)\Gamma\(\frac{s-s'}2\)\Gamma\(\frac{s+s'-1}2\)}{\Gamma(s'-\frac12)\zeta(2s'-1)}E_{s'}(\tau) \nn\\&+ \frac{|J|^{\frac12-s}\pi^{\frac12-s}}4\sum_{n=1}^{\infty} \frac{a_J^{(n,+)}\nu^+_n(\tau)}{(\nu^+_n, \nu^+_n)}\Gamma\(\frac{s-iR^+_n-\frac12}{2}\)\Gamma\(\frac{s+iR^+_n-\frac12}{2}\).
\label{eq:specdecour}
\end{align} 

In the math literature it is more conventional to study a different Poincar\'e series involving a seed with nonzero spin that grows exponentially at the cusp. These are sometimes known as Niebur-Poincar\'e series, and they are defined as \cite{nmj/1118794882}
\begin{equation}
    F(s,\tau,J) \coloneqq \sum_{\gamma \in \Gamma_\infty \backslash PSL(2,\mathbb{Z})} e^{2\pi i J x}\sqrt{y}I_{s-\half}(2\pi |J| y).
\end{equation}
These are convenient objects to study because they are eigenfunctions of the Laplacian
\begin{equation}
    \Delta_\tau F(s,\tau,J) = s(1-s) F(s,\tau,J).
\end{equation}
The Fourier decomposition of these modular functions, and the relationship between them and Poincar\'e series that are more directly related to the modular completion of the characters of light states with spin, is well-known (see in particular \cite{Angelantonj:2012gw}). 

\section{Decompactification limit}
\label{sec:decomp}

In this appendix we review and slightly generalize the derivation in \cite{Angelantonj:2011br} of the $d$-dimensional Eisenstein series in the limit that some of the directions are taken to be large.  More specifically, we consider the limit where the Narain lattice factorizes into 
\be
\Gamma_d = \Gamma_D \times \Gamma_{d-D}.
\ee
and we take the limit where the metric $g^{(D)}_{ij}$ on the $D$-dimensional subspace is $g_{i j}^{(D)}= r^2 \hat{g}_{ij}^{(D)}$ with $r\rightarrow \infty$ and $\det (\hat{g}) \ne 0$.  After Poisson resummation, the lattice sum for $\Gamma_D$ can be written
\be
\Gamma_D =r^{D} \sqrt{\det \hat{g}^{(D)}} \sum_{m^i, n^i} \exp\left(  - r^2 \pi \frac{(m^i + n^i \tau) \hat{g}^{(D)}_{ij} (m^j + n^j \bar{\tau})}{\tau_2} + 2 \pi i B_{ij}m^i n^j\right).
\label{eq:PoissonNarain}
\ee
At large $r$, each term in the sum in $\Gamma_D$ is exponentially suppressed due to the first term in the exponent, unless $n^i=0$ and $\tau_2 \gtrsim r^2$.\footnote{When $\vec{m}= \alpha \vec{n}$ for some $\alpha \in (-\frac{1}{2}, \frac{1}{2})$, there is also an unsuppressed region $\tau_2 \lesssim r^{-2}, (\tau_1 - \alpha) \lesssim r^{-4}$, but this is not contained in the fundamental domain when $r$ is large.} Therefore, up to exponentially small (in $r$) corrections, we can set $n^i=0$ and just work with the sum on $m^i$; this is a significant simplification compared to the original (i.e. not Poisson resummed) expression for $\Gamma_D$, where we would have to include terms both from the sum over momentum modes and winding modes to capture all the power-law terms in $r$.  The basic idea of the decompactification calculation in \cite{Angelantonj:2011br} is simply to combine this Poisson resummation with Zagier's modification of the Rankin-Selberg transform, which tells us exactly how to deal with the non-convergence of the $dy$ integral due to the growth at large $y$ of the vacuum state.
  Schematically, the important structural aspects of the calculation are that after Poisson resummation, $\Gamma_D$  decomposes into the following three pieces:
\be
\Gamma_D = \tilde{\Gamma}_{n=0 \atop m=0} + \tilde{\Gamma}_{n=0 \atop m\ne 0} + \tilde{\Gamma}_{n\ne 0},
\ee
where the subscripts indicate which terms in the sum from (\ref{eq:PoissonNarain}) are included.
The first piece is just a constant, $\tilde{\Gamma}_{n=0 \atop m=0} =\sqrt{\det g^{(D)}} $. The third piece $\tilde{\Gamma}_{n\ne 0}$ is exponentially suppressed at large $r$ and can be discarded.  The second piece $\tilde{\Gamma}_{n=0 \atop m\ne 0}$ vanishes at large $r$ {\it except} for the region where $y \gtrsim r^2$, so anywhere it appears in the integrand we can take $y$ to be large.  In particular, when $\tilde{\Gamma}_{n=0 \atop m\ne 0}$ appears in the integrand, we can exchange the cut-off fundamental domain ${\mathcal F}_T$ for the cut-off strip without modifying the power-law-in-$r$ pieces.

To start, write the RS transform (\ref{eq:ZRS}) with $f = \Gamma_D \times \Gamma_{d-D}$ in the following form:
\begin{eqnarray}
\Lambda(s) R_s[f] &=& \sqrt{\det g^{(D)}}  \left( \int_{\mathcal{F}_T} \frac{dx dy}{y^2} \Gamma_{d-D} E^\star_s + \int_{\mathcal{F}-\mathcal{F}_T} \frac{dx dy}{y^2} \Gamma_{d-D} E^\star_s \right) \nonumber \\
&& +\int_{\mathcal{F}_T} \frac{dx dy}{y^2}   \tilde{\Gamma}_{n=0 \atop m\ne 0}  \Gamma_{d-D} E^\star_s
 + \int_{\mathcal{F}-\mathcal{F}_T} \frac{dx dy}{y^2}   \tilde{\Gamma}_{n=0 \atop m\ne 0}  \Gamma_{d-D} E^\star_s  \nonumber\\
 && +\int_{\mathcal{F}_T} \frac{dx dy}{y^2}   \tilde{\Gamma}_{ n\ne 0}  \Gamma_{d-D} E^\star_s
 + \int_{\mathcal{F}-\mathcal{F}_T} \frac{dx dy}{y^2}   \tilde{\Gamma}_{n \ne 0}  \Gamma_{d-D} E^\star_s  \nonumber\\
&& -  \int_{\mathcal{F}-\mathcal{F}_T} \frac{dx dy}{y^2}\varphi(y) \varphi^\star_s(y) - \Lambda(s) h_s^{(T)} - \Lambda(1-s) h_{1-s}^{(T)},
\end{eqnarray}
where $E_s^\star \coloneqq \Lambda(s) E_s, \varphi^\star_s \coloneqq \Lambda(s) \varphi_s$ and $\varphi(y) = y^{d/2}$. At large $r$, the third line is exponentially suppressed and can be discarded. Because $\varphi(y)$ is a pure power, the terms on the last line above sum to zero.  The  first line is almost $\sqrt{\det g^{(D)}}$ times the RS transform of $ \Gamma_{d-D}$, but is missing the last few pieces from (\ref{eq:ZRS}).  We can add and subtract these pieces to get

\begin{eqnarray}
\Lambda(s) R_s[f] &\cong&  \sqrt{\det g^{(D)}} \Lambda(s)R_s[\Gamma_{d-D}]  \\
&& +\int_{\mathcal{F}_T} \frac{dx dy}{y^2}   \tilde{\Gamma}_{n=0 \atop m\ne 0}  \Gamma_{d-D} E^\star_s
 + \int_{\mathcal{F}-\mathcal{F}_T} \frac{dx dy}{y^2}   \tilde{\Gamma}_{n=0 \atop m\ne 0}  \Gamma_{d-D} E^\star_s  \nonumber\\
&& + \sqrt{\det g^{(D)}} \left( \int_{\mathcal{F}-\mathcal{F}_T} \frac{dx dy}{y^2}   y^{\frac{d-D}{2}} \varphi_s^\star + \Lambda(s) \int_0^T dy y^{\frac{d-D}{2}} y^{s-2} + \Lambda(1-s)\int_0^T dy y^{\frac{d-D}{2}} y^{-1-s} \right).  \nonumber 
\end{eqnarray}
As we discussed above, anywhere that $ \tilde{\Gamma}_{n=0 \atop m\ne 0} $ appears inside the integrand, the only contributions that are not exponentially suppressed come from large $y \gtrsim r^2$, in the second line we can  keep only the zero mode pieces of $\Gamma_{d-D} \sim y^{\frac{d-D}{2}} $ and $E_s^\star\sim \varphi_s^\star$.  
Moreover, because all the integrands in the last line are pure powers of $y$, we can exchanges the integration regions $y \in (0,T)$ with $y \in (T,\infty)$ by adding an overall minus sign.  By choosing $s$ appropriately in various terms in the integrand, we can make all the integrals over $y \in (T,\infty)$ converge and vanish at $T\rightarrow \infty$, so the only terms that survive at $T\rightarrow \infty$ are
\begin{eqnarray}
\Lambda(s) R_s[f] &\cong& \sqrt{\det g} \Lambda(s)R_s[\Gamma_{d-D}] \\
&& +  \sqrt{\det g} \int_0^\infty \frac{dy}{y^2} \sum_{m \in \mathbb{Z}^D/\{0\}} e^{- \frac{\pi m^i g_{ij}^{(D)} m^j}{y}} \left( \Lambda(s) y^{s+\frac{d-D}{2}} + \Lambda(1-s) y^{1-s+\frac{d-D}{2}}\right). \nonumber 
\end{eqnarray}
The remaining integral over $y$ can be done term-by-term in the sum over $m^i$.  To go farther, we can restrict to the cases $D=1$ and $D=2$, where we can do the sums over $m^i$ in terms of known functions.  For $D=1$, the sum over $m$ can be expressed simply in terms of the zeta function, and for $D=2$ in terms of Eisenstein series $E_s$. The case $D=1$ was already obtained by this method in \cite{Angelantonj:2011br}, with the result
\be
\Lambda(s) R_s[f] \cong R_1 \Lambda(s)R_s[\Gamma_{d-1}] + 2 \Lambda(s) \Lambda(s+\frac{d-1}{2}) R_1^{2s+d-2} + 2 \Lambda(1-s) \Lambda(1-s+\frac{d-1}{2}) R_1^{d-2s}.
\ee
where $R_1 \coloneqq \det g$.  For $D=2$, we can use the fact that
\begin{equation}
 \sum_{m\in\mathbb{Z}^2/\{0\}}(m^i g_{ij}^{(2)}m^j)^{-s} =  \sum_{m\in\mathbb{Z}^2/\{0\}} \left({\sigma_2\over \rho_2|m_1+\sigma m_2|^2}\right)^s = 2\rho_2^{-s}\zeta(2s)E_s(\sigma)
\end{equation}
to obtain
\be
\Lambda(s)R_s[f] \cong \rho_2 \Lambda(s) R_s[\Gamma_{d-2}] + 2 \rho_2 \left( \Lambda(s) \rho_2^{-s+\frac{d-2}{2}} E_{s-\frac{d-2}{2}}^\star(\sigma) + \Lambda(1-s) \rho_2^{s-1+\frac{d-2}{2}} E_{1-s-\frac{d-2}{2}}^\star(\sigma) \right) .
\ee
Finally, consider specifically the case of the $T^2\times T^2$ locus of the $c=4$ moduli space, where the volume of one of the tori is taken to be large, so $d=4$ and $D=2$.  The constant term inherited from this RS transform is then given by
\begin{equation}
	\Res_{s=1}\left(R_{1-s}[Z^{(c=4)}]E_s(\tau)\right) = \Res_{s=1} R_s[Z^{(c=4)}] = \rho_2^{(1)}\left(\widehat E_1(\sigma^{(1)})+\widehat E_1(\rho^{(2)})+\widehat E_1(\sigma^{(2)})+{3\over \pi}\log\rho_2^{(1)} + \gamma\right)
\end{equation}
where $\gamma$ is defined as in (\ref{eq:gammadeff}).
We have checked numerically that at large $\rho_2^{(1)}$, the coefficients of the cusp forms follows the same pattern as we saw at smaller $c$, giving (\ref{eq:c4wooo}). 

\bibliographystyle{JHEP}
\bibliography{harmonicAnalysisbib_v3}
\end{document}